\documentclass[review]{elsarticle}
\usepackage{url,graphicx,epsfig,graphics,subfigure,psfrag,amsmath,amssymb,epstopdf,enumerate,lineno,hyperref}
\modulolinenumbers[5]
\usepackage{diagbox}
\usepackage{xcolor}
\usepackage[T1]{fontenc}    
\usepackage[utf8]{inputenc} 
\usepackage[normalem]{ulem}
\newcommand{\stkout}[1]{\ifmmode\text{\sout{\ensuremath{#1}}}\else\sout{#1}\fi}

\journal{Physica A-Statistical Mechanics and Its Applications}

\begin{document}

\begin{frontmatter}

\title{Learning phase transitions by siamese neural network}
%
\author[mymainaddress,secondaryaddress]{Jianmin Shen}
\author[secondaryaddress,thirdaddress]{Shiyang Chen\corref{mycorrespondingauthor}}
\cortext[mycorrespondingauthor]{Corresponding author}
\ead{SY.Chern@Swansea.ac.uk}

\author[secondaryaddress,fourthaddress]{Feiyi Liu\corref{mycorrespondingauthor}}
\cortext[mycorrespondingauthor]{Corresponding author}
\ead{fyliu@mails.ccnu.edu.com}

\author[secondaryaddress,fifthaddress]{Wei Li}

\author[mymainaddress]{Youju Liu}

\address[mymainaddress]{School of Engineering and Technology, Baoshan University, Baoshan 678000, China}

\address[secondaryaddress]{Key Laboratory of Quark and Lepton Physics (MOE) and Institute of Particle Physics, Central China Normal University, Wuhan 430079, China}

\address[thirdaddress]{Department of Physics, Sweansea university, SA2 8PP, Swansea, United Kingdom}

\address[fourthaddress]{Institute for Physics, E{\"o}tv{\"o}s Lor\'and University\\1/A P\'azm\'any P. S\'et\'any, H-1117, Budapest, Hungary}

\address[fifthaddress]{SCIQ Lab, École Supérieure d'Informatique Électronique Automatique, Ivry-sur-Seine 94200, France}

\begin{abstract}
The wide application of machine learning (ML) techniques in statistics physics has presented new avenues for research in this field. In this paper, we introduce a semi-supervised learning method based on Siamese Neural Networks (SNN), trying to explore the potential of neural network (NN) in the study of critical behaviors beyond the approaches of supervised and unsupervised learning. By focusing on the (1+1) dimensional bond directed percolation (DP) model of nonequilibrium phase transition and the 2 dimensional Ising model of equilibrium phase transition, we use the SNN to predict the critical values and critical exponents of the systems. Different from traditional ML methods, the input of SNN is a set of configuration data pairs and the output prediction is similarity, which prompts to find an anchor point of data for pair comparison during the test. In our study, during test we set different bond probability $p$ or temperature $T$ as anchors, and discuss the impact of the configurations at this anchors on predictions. In addition, we use an iterative method to find the optimal training interval to make the algorithm more efficient, and the prediction results are comparable to other ML methods.
\end{abstract}


\end{frontmatter}


\section{Introduction}
\label{intro}
With the rapid progress of the computer field in the past decade, machine learning (ML) has been explosively applied in various fields of science~\cite{jordan2015machine,mohri2018foundations}.   By the great power of feature capture and classification, ML has brought new inspiration to the study of statistical physics and proven to be an effective tool for identifying and classifying different phases~\cite{mehta2014exact,carleo2017solving,carrasquilla2017machine,wang2016discovering,van2017learning}. Through training process, ML algorithms can learn the underlying patterns and laws of phase transition from the input configurations of a given system, enabling them to make predictions in test data. Additionally, ML algorithms can automatically extract and select the most relevant features, which are valuable for reducing the complexity of high-dimensional phase transition data.

The mainly types of ML method applied to study phase transitions are supervised learning~\cite{carrasquilla2017machine,van2018learning,canabarro2019unveiling,ni2019machine,wang2023supervised,bayo2023percolating}, unsupervised learning~\cite{wang2016discovering,PhysRevE.96.022140,hu2017discovering,wang2021unsupervised,kaming2021unsupervised,Shen.SciRep.2022,wang2023supervised,qi2025principal}, and semi-supervised learning ~\cite{el2016asymptotic,bencteux2020automatic,court2018auto,shen2022transfer,CHEN2023128666}. 
For supervised learning, the input data for training needs to be labeled, which commonly used to identify or classify the phases of matter. For unsupervised learning, the labeling of the order parameter is not required, such as the methods of principal component analysis (PCA)\cite{wang2016discovering, PhysRevE.96.022140,Shen.SciRep.2022,qi2025principal}, t-distributed stochastic neighbor embedding (T-SNE)~\cite{PhysRevE.97.013306} or nonlinear autoencoder (AE)~\cite{PhysRevE.97.013306,wang2023supervised,PhysRevE.96.022140,Shen.SciRep.2022,ghosh2022supervised}, for clustering and dimensionality reduction. 
Recently, semi-supervised learning, especially transfer learning (TL) also has been proved as a good choice to detect critical points and calculate critical exponents~\cite{shen2022transfer}, even determining the type of phase transition~\cite{CHEN2023128666}.
With the input of mixing both labeled and unlabeled data, methods of TL can transform unlabeled data in target domain into labeled data in source domain by training, which only use a limited labeled set for the purposes of prediction. 
Although the methods of ML are dazzling, all of them are not fully perfect. For example, supervised learning must time-consumingly label the data; unsupervised learning is hard to fully identify the dynamic features with limited data; semi-supervised learning is very sensitive to the input, which always need to be preprocessed. Therefore, it is still imperative to find a more complete and efficient method of ML to study phase transitions.

Siamese neural network (SNN) \cite{chicco2021siamese,ranasinghe2019semantic,zhang2016siamese}, also called twin NN, is a specialized architecture usually consisting of two parallel identical neural networks (NN) that share their weight for the purposes of assessing or comparing the similarity between two input objects. SNN is not simply classified into supervised learning, unsupervised learning or semi-supervised learning, but a flexible structure that fill them in. It is widely used in computer vision for image recognition~\cite{wu2017face}, target tracking~\cite{bertinetto2016fully} and image retrieval~\cite{qi2016sketch}, and also shows its power in natural language processing (NLP) for text matching tasks~\cite{ichida2018measuring}. In the study of phase transition, Patel $et al$~\cite{patel2022unsupervised} introduced a SNN technique of unsupervised learning to identify phase boundaries in Ising-type systems and Rydberg atomic arrays, showing its ability to learn about multiple phases without knowing about their existence. The SNN of convolutional neural network(CNN) architectures has also been used to investigate phase transitions based on polarising microscopic textures of liquid crystals, showing high accuracy for phase distinguishing~\cite{osiecka2024siamese}. As the preliminary attempt, SNN is proved as a valuable tool for discovering new and unfamiliar phases of matter, and worth for further application and exploration. Unlike general ML methods used on phase transition models, the SNN is not intended to predict which phase an input configuration belongs to, but judges whether different configurations belong to a same phase. Therefore, the output of the SNN is a measure of similarity, not a label associated with the input configuration. The SNN compares the configuration of unknown labels with the configuration of specific anchor points, transforming the process of assigning predicted labels into judging similarity. This means the selection of training and test intervals becomes very flexible, allowing SNNs to handle not only two-phase but also multi-phase complex statistical models~\cite{ccivitciouglu2025phase}, with a significant improvement in computing efficiency.

Inspired by this, we apply a new SNN algorithm based on semi-supervised learning to study the critical behaviors of phase transition. In this method, we divide input configurations to two separate sets for training. By choosing an anchor for configurations, the SNN would determine whether this samples belong to the same set through training, which gives the power to judge phases and find critical point. For test, we focus on two key models. The first is the (1+1) dimensional bond-directed percolation (DP)~\cite{obukhov1980problem,grimmett2002directed,henkel2008non}, which belongs to the DP universality class, a prominent class of non-equilibrium phase transitions~\cite{doi:10.1080/00018730050198152,RevModPhys.76.663,doi:10.1142/S0217979204027748}. The second model is the 2 dimensional Ising model~\cite{onsager1944crystal,lee1952statistical}, which belongs to the Ising universality class. We not only predict the critical point of the DP model through the SNN, but also calculate the critical exponent by the data collapse method~\cite{bhattacharjee2001measure,kimchi2018scaling}. Further, we discuss 
the impact of anchor point selection on results, and find the optimal interval of training set to achieve more accurate predictions. By doing so, we hope to enhance the effectiveness and reliability of the SNN as a new tool in the study of phase transitions. 

The remaining content of this paper is structured in the following manner. In Section \ref{models}, we have introduced the two models. Section \ref{method} gives the methods of SNN and the data set for study. Section \ref{sec:Results} shows the estimate critical probabilities of SNN and also the prediction of spatial correlation exponent $\nu_\perp$. In Section \ref{discussion}, we discuss the case of chosen archor in critical region, and the impact of the training set on the results. Section \ref{conclusion} is a summary of this work.

\section{Models}
\label{models}
\subsection{The directed percolation model}
\label{directedpercolation}

As the most prominent universality class of absorbing phase transition, DP describes percolation of one direction along that the system falls into inactive state where the time evolution stops.
It is used to investigate various phenomena, including the propagation of epidemics~\cite{grassberger1983critical}, forest fires~\cite{von1986dynamics}, and traffic flow ~\cite{nagatani1993jamming}, etc. In past decades, various methods have been applied to study DP models, such as mean-field theory~\cite{de1983directed}, renormalization groups~\cite{kinzel1981directed}, field theory approach~\cite{janssen2005field} and numerical simulations~\cite{hinrichsen1998numerical} and so on. Some ML approaches have also been tried recently and shown their own strengths~\cite{shen2022transfer,PhysRevE.103.052140}. Since the models of DP universality class include unitary or binary random reaction processes, which may contain diffusion or non-diffusion motions, only the critical exponents are common and the position of the phase transition may vary. In our work, we use a SNN based algorithm to study the $(1+1)$-dimensional bond DP, one of the simplest cases of DP universality class. 

\begin{figure}[tbp]
\centering
\subfigure[$p<p_c$]{
\label{Fig.05}
\includegraphics[height=0.29\textwidth]{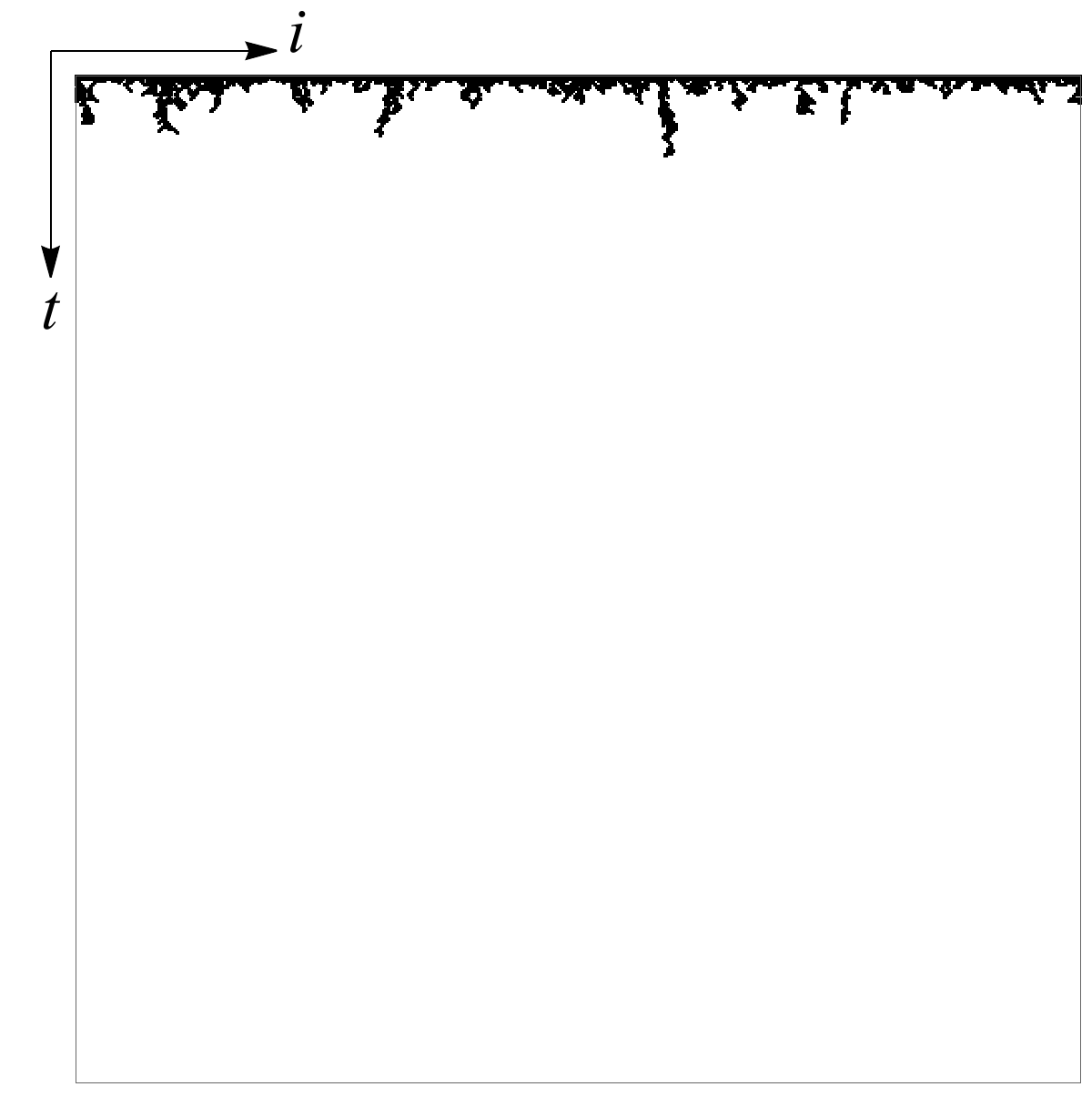}}
\subfigure[$p=p_c$]{
\label{Fig.6447}
\includegraphics[height=0.29\textwidth]{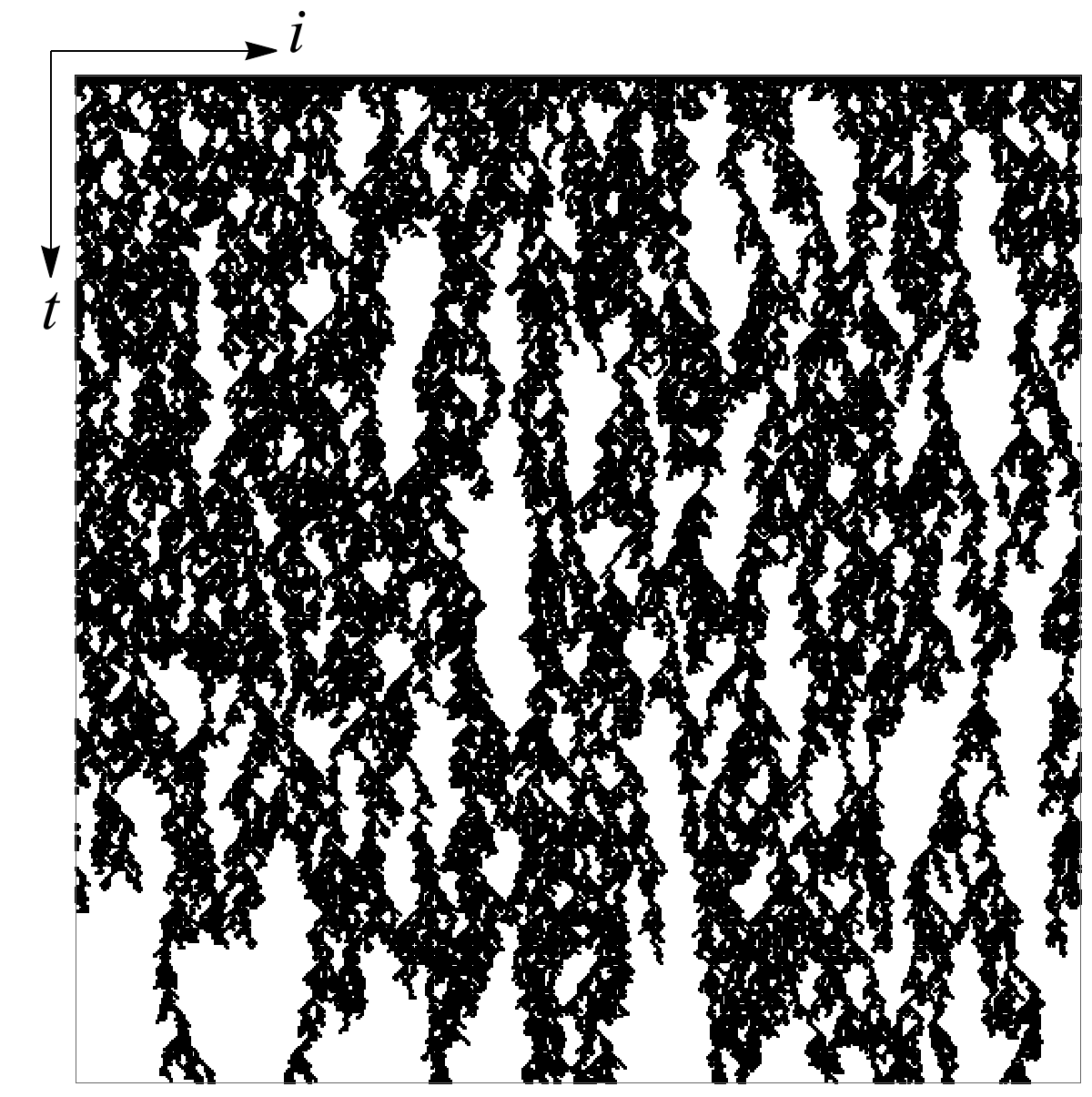}}
\subfigure[$p>p_c$]{
\label{Fig.08}
\includegraphics[height=0.29\textwidth]{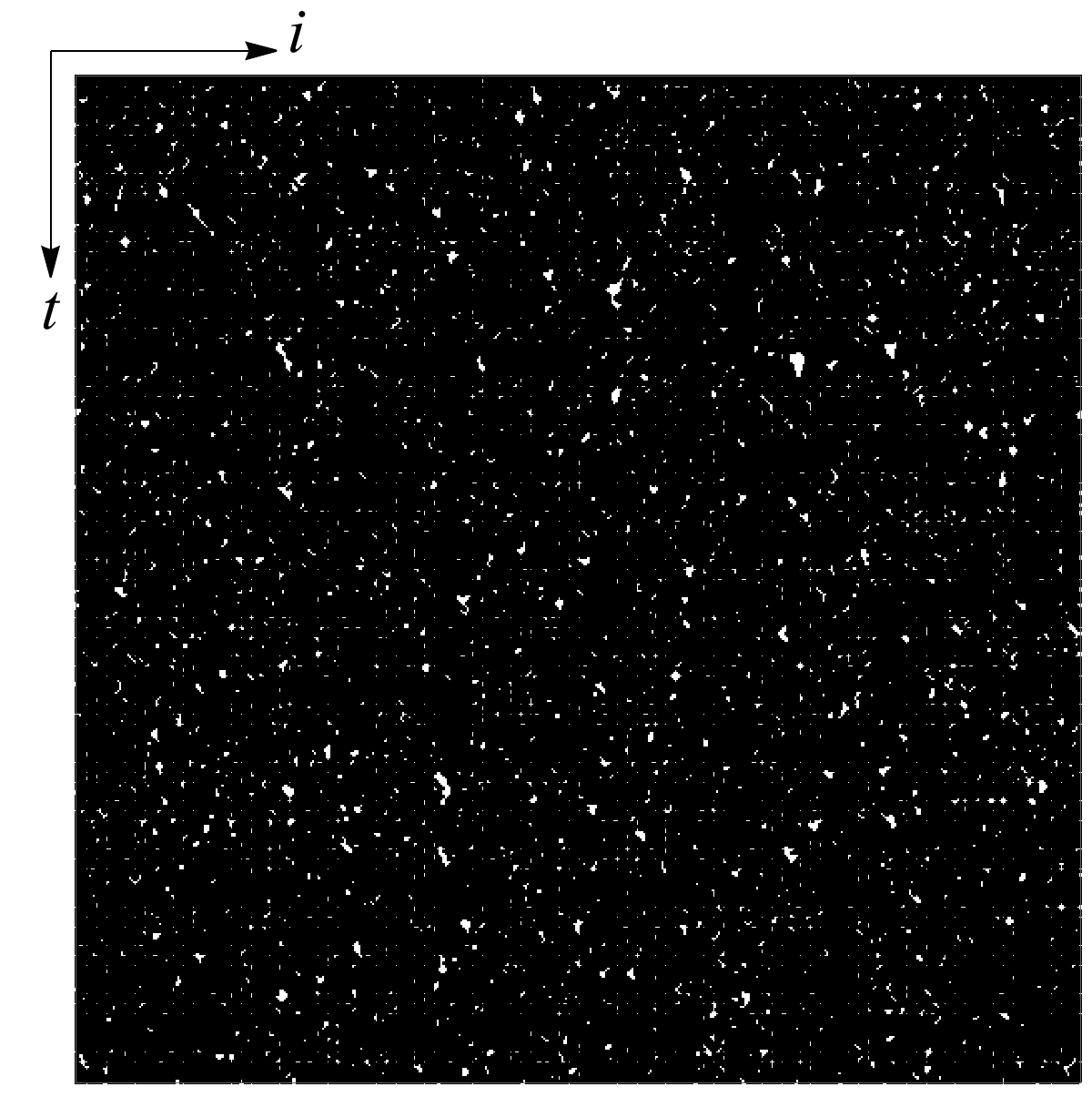}}
\caption{Configurations of the (1+1) dimensional bond DP corresponding to three different bond probabilities $p=0.5,0.6447$ and $0.8$ respectively. The Black sites represent occupied sites, while the white areas indicate empty regions. The system configuration corresponds to $L=500$ with $t=500$ time steps. The vertical and horizontal arrows denote the temporal evolution direction and spatial dimensions, respectively. The periodic boundary conditions are implemented along the spatial axis.}
\label{configuaration_dp}
\end{figure}

Fig.~\ref{configuaration_dp} illustrates the configurations of (1+1)-dimensional DP on the square lattice of size $L=500$ with the evolution time step $t=500$. Starting from a fully occupied lattice, a bond is formed at the time step with probability $p$ from an existing bond, which can be interpreted as a reaction-diffusion mechanism of interacting particles with the active particle $A$ and the empty site $\varnothing$:
\begin{equation}
\label{DP-diff}
  \left\{
\begin{array}{rcl}
 \mbox{self-destruction:} &   A \longrightarrow \varnothing,  \\
 \mbox{diffusion:} &   \varnothing + A \longrightarrow A + \varnothing, \\
 \mbox{offspring production:} &  A \longrightarrow 2A, \\
 \mbox{coagulation:} & 2A \longrightarrow A. \\ 
\end{array} \right.
\end{equation}

By the arrows, this system evolves with the following rules:
\begin{equation}
  s_{i}(t+1)=\left\{
\begin{array}{rcl}
1     &      & {if \quad s_{i-1}(t) = 1 \quad \text{and} \quad z_{i}^{- } < p,} \\
1     &      & {if \quad s_{i+1}(t) = 1 \quad \text{and} \quad z_{i}^{+ } < p,} \\
0     &      & \text{otherwise},
\end{array} \right.
\end{equation}
where $s_{i}(t)$ is the state of site $i$ at time $t$ and $p$ is the bond probability of a connection between two adjacent sites. $z_{i}^{\pm} \in (0, 1) $ is a random number taken from a uniform distribution. The system employs periodic boundary conditions along the spatial axis, as illustrated in Fig. \ref{configuaration_dp}. Each active site $s_{i}(t)$ propagates simultaneously to its left $s_{i-1}(t)$ and right $s_{i+1}(t)$ neighboring spatial positions during temporal evolution. The mutual independence of propagation probabilities $z_{i}^{\pm} \in (0, 1)$ ensures the absence of spatial correlations. From the definitions above, the configuration of the system can be determined by the states of all sites.

The models of DP universality class also have their own independent critical exponents. The order parameter of the (1+1)-dimensional bond DP close to the critical point $p_c$ can be expressed by the steady-state particle density $\rho_{a}$ according to the critical exponent $\beta$ as~\cite{doi:10.1080/00018730050198152},
\begin{equation}
 \rho_{a}(p)\widetilde{\propto}(p-p_{c})^{\beta}.
\end{equation}
In this steady state, the system also obeys the following power-law expressions of spatial and temporal correlation length:
\begin{equation}
  \xi_{\perp}\sim \mid p-p_{c}\vert^{-\nu_{\perp}},   \quad  \xi_{\parallel}\sim \mid p-p_{c}\vert^{-\nu_{\parallel}},
\end{equation}
where $\nu_{\perp}$ and $\nu_{\parallel}$ are the spatial and temporal correlation exponent.

According to established literature, the theoretical critical threshold for (1+1)-dimensional bond percolation is $p_c=0.644700185(5)$~\cite{jensen1999low}, determined through series analysis. Recent Monte Carlo simulations with high precision implementation yield $p_c=0.6447001(2)$~\cite{wang2013high}, obtained by single particle sources with $t=2^{14}$ time steps.

In the (1+1)-dimensional bond directed percolation (DP) model, there are two types of initial conditions for generating configurations. One is the homogeneous particle source, i.e., a fully occupied lattice where all spatial sites are initially occupied; the other is the single particle source, as only one site is initially occupied. Due to the reflection symmetry of the DP dynamics, both initial conditions lead to the same critical behavior, including the critical point and the critical exponents. In this work, we use the fully occupied lattice, as shown in Fig.~\ref{configuaration_dp} as input to the SNN algorithm. In the generated configurations, the first row represents the system at time $t=0$. Each row corresponds to spatial information at a given time step, and each column reflects the temporal evolution. To generate the configurations, we use periodic boundary conditions and originally start with fully occupied lattice (explained in our previous study~\cite{shen2022transfer}). The characteristic temporal length $t_c\sim L^{z/d}$ is set by the dynamical exponent $z=1.580(1)$ and the spatial dimension $d=1$, to ensure stable results~\cite{doi:10.1080/00018730050198152}.

\subsection{The Ising model}
\label{isingmodel}

\begin{figure}[tbp]
\centering
\subfigure[$T<T_c$]{
\label{Fig.152}
\includegraphics[height=0.23\textwidth]{ising152.pdf}}
\subfigure[$T=T_c$]{
\label{Fig.2269}
\includegraphics[height=0.23\textwidth]{ising2269.pdf}}
\subfigure[$T>T_c$]{
\label{Fig.352}
\includegraphics[height=0.23\textwidth]{ising352.pdf}}
\caption{Configurations of the 2 dimensional Ising model corresponding to three different temperatures $T=1.52,2.269$ and $3.52$ respectively. The red lattice points represent spin-up states$(\uparrow)$ and the blue lattice sites indicate spin-down states$(\downarrow)$. The system configuration corresponds to a lattice size of L=32 with t=10000 time steps.}
\label{configuaration_ising}
\end{figure}

The Ising model~\cite{onsager1944crystal,lee1952statistical} serves as a paradigmatic model for systems exhibiting strong interactions between particles. Within the phenomenological framework, phase transitions can be characterized by order parameters, which assume non-zero values in ordered phases and vanish in disordered phases~\cite{landau1966lehrbuch}. The two-dimensional Ising model considers a periodic lattice with exchange interactions limited to nearest neighbors. In the case of uniaxial magnetic materials, each lattice site can adopt either a spin-up or spin-down state. The Hamiltonian of the classical two-dimensional Ising model can be expressed as:
\begin{equation}
H = -J \sum \limits_{<i,j>} \sigma_{i} \sigma_{j} - \sum \limits_{j}h_{j} \sigma_{j},
\end{equation}
where $\sigma_{i} \in \left\{-1,1 \right\}$ represents the classical spin, $J$ denotes the interaction strength, and $h_{j}$ represents the external field. The critical temperature at which this model undergoes a phase transition is $T_{c} = 2.269$ (for $h_{j} = 0$)\cite{onsager1944crystal}. Fig.~\ref{configuaration_ising} depicts the configurations of the Ising model at three different temperatures. The selection of the ising model dataset for our study will be given in Section~\ref{dataset}.

\section{Method}\label{method}
\subsection{The Siamese Neural Network (SNN) method}\label{snn}
The SNN is a type of deep learning model commonly used for tasks involving metric learning and similarity comparison~\cite{chicco2021siamese}. In various applications, SNNs are frequently employed to analyze pairs of input data, such as images, text, or voice~\cite{chicco2021siamese}. It works by creating multiple parallel sub-networks, known as "Branches", which share the same structure and weights. By feeding each input data through the NN of these branches for feature extraction, the SNN can calculate the similarity scores between them using metric functions like Euclidean distance or cosine similarity. The significant advantage of SNN is the capability to acquire adaptable feature representations while training, which enables the NN to achieve high performance even on datasets having some degree of variability. Additionally, SNN can be trained end-to-end by back-propagation algorithms, which allows the entire NN to autonomously learn feature representations and measures of similarity.

Compared to other ML methods, the SNN employs a dual-branch architecture specifically designed to learn the dynamic similarity between pairs of configurations, making it particularly effective at capturing spatiotemporal correlations. It requires only “same-phase/different-phase” labels for configuration pairs—no explicit phase or critical-point labels for individual samples—resulting in much more efficient labeling, especially for ambiguous configurations near the critical region. Moreover, the same SNN architecture can be applied without modification to both equilibrium systems (e.g., the Ising model) and non-equilibrium systems (e.g., DP), demonstrating strong cross-system generalizability.

However, SNN also has its limitations. It outputs a similarity score rather than explicit class probabilities, which can make interpretation more challenging—an issue that has been noted in reviews of contrastive learning approaches. Its performance is highly sensitive to how representative the positive and negative sample pairs are; if these pairs are imbalanced or poorly chosen, training can suffer. Moreover, the deep and abstract nature of the learned feature space means that SNN embeddings often do not directly map to measurable physical quantities. Overall, while not a replacement for traditional or other ML methods, SNN offers a powerful complementary tool—especially for analyzing systems with complex spatiotemporal dynamics.

\begin{figure}[htbp]
\centering
\includegraphics[width=0.9\textwidth]{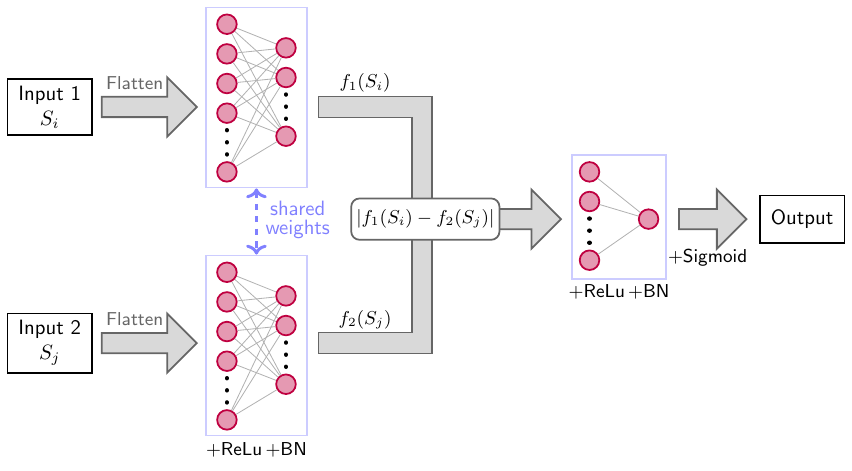}
\caption{The structure of SNN. \( S_i \) and \( S_j \) represent a pair of input configurations (e.g., spatiotemporal configurations generated by bond probabilities \( p_i, p_j \) in the DP model, or spin configurations generated by temperatures \( T_i, T_j \) in the Ising model).  
\texttt{Flatten} denotes the operation of flattening a two-dimensional configuration into a one-dimensional vector.  
The sequence \( \rightarrow \mathrm{BN} \rightarrow \mathrm{ReLU} \) indicates that the fully connected layer is followed by Batch Normalization and Rectified Linear Unit (ReLU) activation.  
The term \( \|f_{1}(S_i) - f_{2}(S_j)\| \) denotes the feature difference between the two inputs. This difference is passed through Batch Normalization and ReLU activation, and then a \texttt{Sigmoid} activation function is applied to produce a similarity probability output.}
\label{fig:SNN}
\end{figure}

The specific SNN framework used in our work to study the phase transition of the (1 + 1) dimensional DP is shown in Fig.~\ref{fig:SNN}. For the better understanding, we name the first part of the SNN which produces $f(S_i)$ as Feature extractor, and the later one estimates similarity as Similarity estimator. 
We use the raw configurations of $L =8, 16, 32, 48$ and $64$ as input. The input is a set of data pairs $(S_i, S_j)$, $S_i$ enters one branch in the framework, and $S_j$ enters the other. 
The branches here are generated with the replica of the Feature extractor, sharing weight and the same hyper-parameters.
$S_i$ and $S_j$ are two different configuration samples of the same type of data. Taking the branch with $S_i$ as an example, the input $S_i$ is flattened into one-dimensional data and fed into Feature extractor. $S_i$ flows firstly into a fully connected (FC) layer of $500$ neurons with a $ReLu$ activation function. Here a Batch Normalization (BN) is set to avoid over-fitting and speed up the NN to optimize its parameters before the $Relu$ activation function. One more FC layer in the Feature extractor maps the data flow into a feature representation $f_1$ of length $50$. $S_i \in \mathbb{R}^{L \times T}$ denote the input configuration pair, where $L$ is lattice size and $T$ is temporal dimension. The feature extraction process can be formalized as:
\begin{equation}
f(S_i) = W_{\text{feat}} \cdot h^{(1)} + b_{\text{feat}}, 
\end{equation}
with $h^{(1)}$ the hidden layer,
\begin{equation}
    h^{(1)} = \text{ReLU}\left(\text{BN}(W_{\text{flat}} \cdot \text{vec}(S_i) + b_{\text{flat}})\right).
\end{equation}
Here $\text{vec}(\cdot)$ means a flatten operation, and $\text{BN}(\cdot)$ gives the Batch Normalization. $W_{\text{flat}} \in \mathbb{R}^{500 \times (L \cdot T)}$ and $W_{\text{feat}} \in \mathbb{R}^{50 \times 500}$ are the first and second FC layer weights, respectively. The entire calculation process of the first branch can be expressed as function $f_1(S_i)$. Similarly, the second branch with $S_j$ has the same process, and the feature representation is recorded as $f_2(S_j)$. Note that these two branches shares the weights for learning.

In the SNN, the weight update in the shared weight area is basically the same as in a single structure NN, except that the shared weights would update the same part in both branches simultaneously. Each branch of the SNN processes one input separately, and the forward propagation of the shared weight area calculates the same weights and biases. After that, the outputs of the two branches are merged to calculate a loss value. Once the loss is calculated, the gradient is determined through backpropagation. The gradient of the shared weight area is the sum of the gradients from the two branches. This shared weight mechanism ensures that the two branches remain consistent during training, thereby achieving effective feature extraction and comparison of similar or different inputs. Since the shared weights are the same in both branches, the weights are updated simultaneously.

The shared weights scheme enable Feature extractor embeds data to the same space from any branches. Therefore, it allows to calculate the ``distance" between two the feature representations $f_1(S_i)$ and $f_2(S_j)$. in order to compare the similarity of $S_i$ and $S_j$, here we define the ``distance" by $L_1$ norm. Then the norm is fed into Similarity estimator. The $L_1$ norm emphasizes substantial differences in the learned representation space, analogous to the role of an order parameter in highlighting critical changes between phases. As a result, the model focuses on macroscopic statistical distinctions, rather than microscopic alignment. The network consists of two dense layer of FC with neurons $50$ and $1$ neurons and $Relu$ and $sigmoid$ activation function. Before the data flows get through the activation of the first layer, the flow is normalized with BN layer. The output of the Similarity estimator, $s(S_i, S_j)$, is converted into a probability of $[0,1]$ due to the $Sigmoid$ activation function,  representing the similarity between $S_i$ and $S_j$. Then the similarity metric is computed as:
\begin{equation}
s(S_i, S_j) = \sigma \bigg( W_{1\times 50} \cdot \text{ReLU}\bigg(\text{BN}(|f(S_i) - f(S_j)|)\bigg) + b_{50}\bigg),
\end{equation}
where $\sigma(\cdot)$ is sigmoid function.

\subsection{Data sets of the SNN}\label{dataset}

For input data of the SNN, we use the full configurations of (1+1) dimension DP (time evolution step as the characteristic temporal length.). In each lattice size $L =8, 16, 32, 48$ and $64$, we select $101$ values of $p\in[0,1]$ uniformly and generate $1000$ samples of configuration at each $p$.
To minimize human intervention during training, we label the samples of configuration far away from the critical regime, i.e., the input configurations $S_i$ and $S_j$ are generated in the range of $p\in[0,0.1]\cup [0.9,1]$. The idea of labeling here is very different from the input of TL in Ref.~\cite{shen2022transfer}. The labels ``1" or ``0" indicates the input data pair $S_i$ and $S_j$ belong to the same or different phases to ensure maximum similarity or dissimilarity between them. 
Specifically, if $S_i$ and $ S_j$ both belong to $[0,0.1]$ (or $[0.9,1]$), the input $(S_i, S_j)$ is labeled as ``1". Or $S_i$ and $ S_j$ are in the range of $[0,0.1]$ and $[0.9,1]$ respectively, the data pair is labeled as ``0". 

For the 2-dimensional Ising model, we employ the Wolff algorithm to generate the configurations. For each temperature point including those near the critical region($T \approx 2.269$) and in the low-temperature regime($T<T_c$), we perform 10000 Wolff cluster updates. The first $50\%$ of these steps (5000 steps) are exclusively used for thermalization, and sampling is conducted only during the remaining 5000 steps to eliminate the influence of the initial state. The temperature $T$ is selected from the range of 1.02 to 4.02, with an interval of 0.01, resulting in a total
of 301 temperature points. For each temperature, 1000 samples are generated. The label set is chosen as $T\in[1.02,1.52]\cup [3.52,4.02]$. For training, the loss function is defined as:
\begin{equation}
    L = \sum\limits_{i=1}^{N}y(S_i, S_j)\log s(S_i, S_j) + \big(1-y(S_i,S_j)\big)\log\big(1-s(S_i,S_j)\big),
 \label{SNN_loss}
\end{equation}
where $s$ is the similarity of ($S_i$, $ S_j$) , and $y$ is the label, ranging from 0 to 1.  The training epoch is set to $5000$, and our work in on python 3.10 and RTX 4090 Gpu platform with tensorflow 2.14.0 library. 

For the process of test, a reference point, called "anchor" point, is selected to evaluate the similarity $s$ or dissimilarity $(1-s)$ of all configurations. The output $s$ by the SNN represents the average similarity between configurations generated at  specific bond probabilities (for DP) or temperatures (for Ising) and anchor $p_a$. For a given pair $(S_i, S_j)$, the SNN computes their similarity score $s(S_i, S_j) \in [0,1]$. When analyzing similarity across parameter ranges, the reported curve is the average over all pairwise comparisons at each parameter value. This approach allows the SNN to capture the statistical similarity between configurations, even when individual samples may appear random due to fluctuations near criticality. By averaging over multiple configurations, the SNN effectively smooths out noise and highlights the underlying phase-dependent patterns. The anchor $p_a$ can be selected arbitrarily within the range of $p\in[0,1]$. By taking the samples of configuration at the anchor $p_a$ as $S_j$ and the samples of the training set as $S_i$, in this way the similarity between the samples at the anchor and the training samples can be estimated through the SNN. And in detail the dependence of SNN results on anchor point will be discussed in Section~\ref{anchor_dependence}. Then the output of SNN, $P_0$, can be expressed as two curves within bond probability $p\in[0,1]$:  the curve $s$ of similarity and $1-s$ of dissimilarity. The intersection of these two curves or $p$ at $P_0=50\%$ represents the critical point of phase transition in the system. 

\section{Results}\label{sec:Results}
\subsection{Test of the SNN framework}

\begin{figure}[htbp]
\centering
\includegraphics[width=0.6\textwidth]{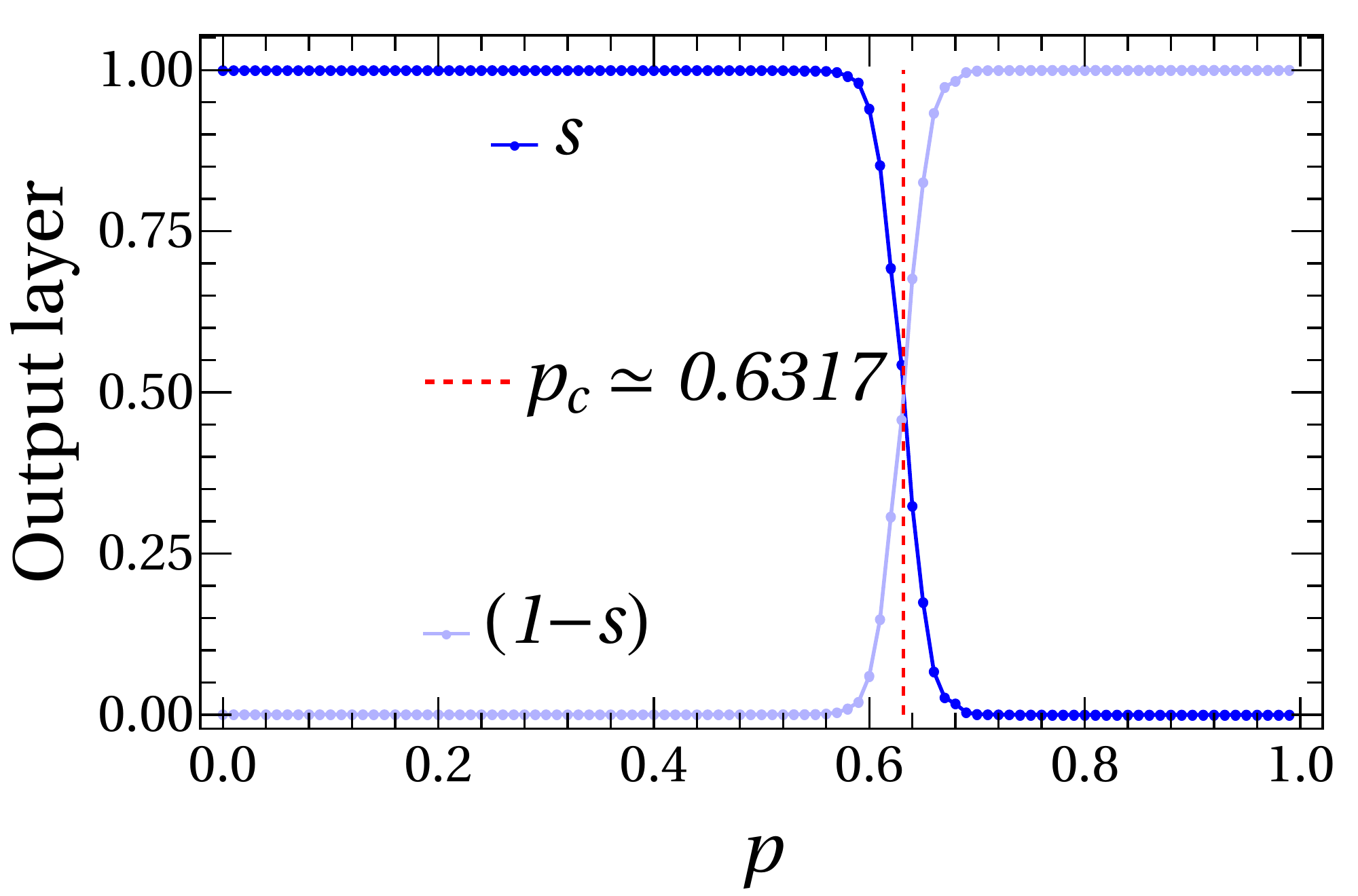}
\caption{The output of SNN with the test anchor $p_{a} = 0$ at $L=32$. The blue curve represents the similarity ($s$), while the ensemble average of the test samples is $1000$. The light blue curve represents the non-similarity ($1-s$).}
\label{fig:32_criticalpoint}
\end{figure}

According to the previous section, the entire SNN framework has been well trained by the set of samples within $p\in[0,0.1]\cup [0.9,1]$. To test the usability of SNN, firstly we select all samples of $p\in[0,0.1]$ as $S_i$ and only the configurations of anchor $p_a=0$ as $S_j$ at $L=32$ to form the input of test. The output $P_0$ is shown in the Fig.~\ref{fig:32_criticalpoint}. For the similar curve $s$, $P_0$ is a stable value $1$ at $p\in[0,0.48]$ and $0$ at $p\in [0.64, 1]$, which means all configurations of anchor $p_a=0$ are $100\%$ having the same phase with samples of $p\in[0,0.48]$ but totally different with the samples of $p\in [0.64, 1]$. In the area of $p\in (0.48,0.64)$, the similarity curve $s$ drops sharply, that is, the samples begin to be dissimilar to the ones of anchor $p_a=0$, and then it is completely in differ near $p=0.64$. The result of dissimilarity curve $(1-s)$ is opposite to the similarity. The evaluated critical point of SNN at $L=32$ is $p_c\simeq0.6317$, which is close to the theoretical value $0.644700185(5)$~\cite{jensen1999low}. This shows that the SNN can relatively accurately predict the critical point of the (1+1) dimensional bond DP at $L=32$ by the selected anchor $p_a=0$ . 
 
\subsection{The dependence of anchor selection}\label{anchor_dependence}

\begin{figure}[htbp]
\centering
\subfigure[$L= 16$]{
\label{Fig.16.}
\includegraphics[height=0.29\textwidth]{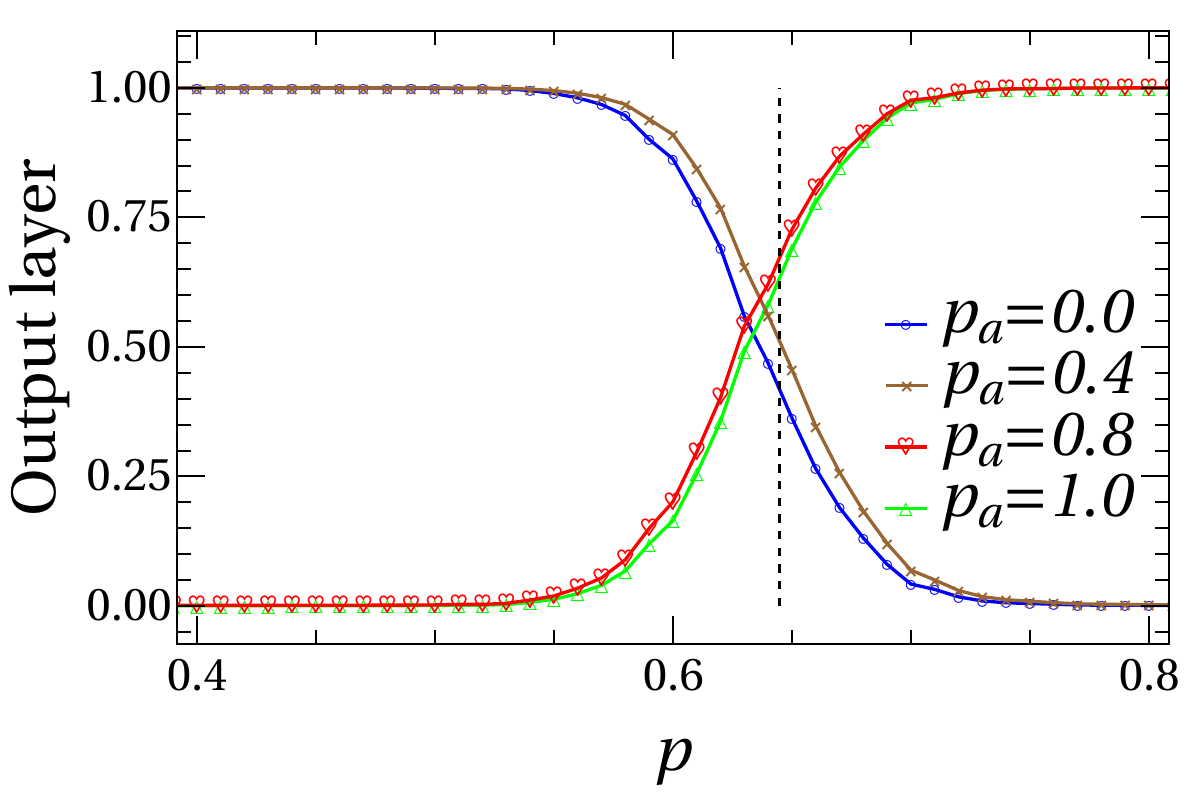}}
\subfigure[$L= 32$]{
\label{Fig.32.}
\includegraphics[height=0.29\textwidth]{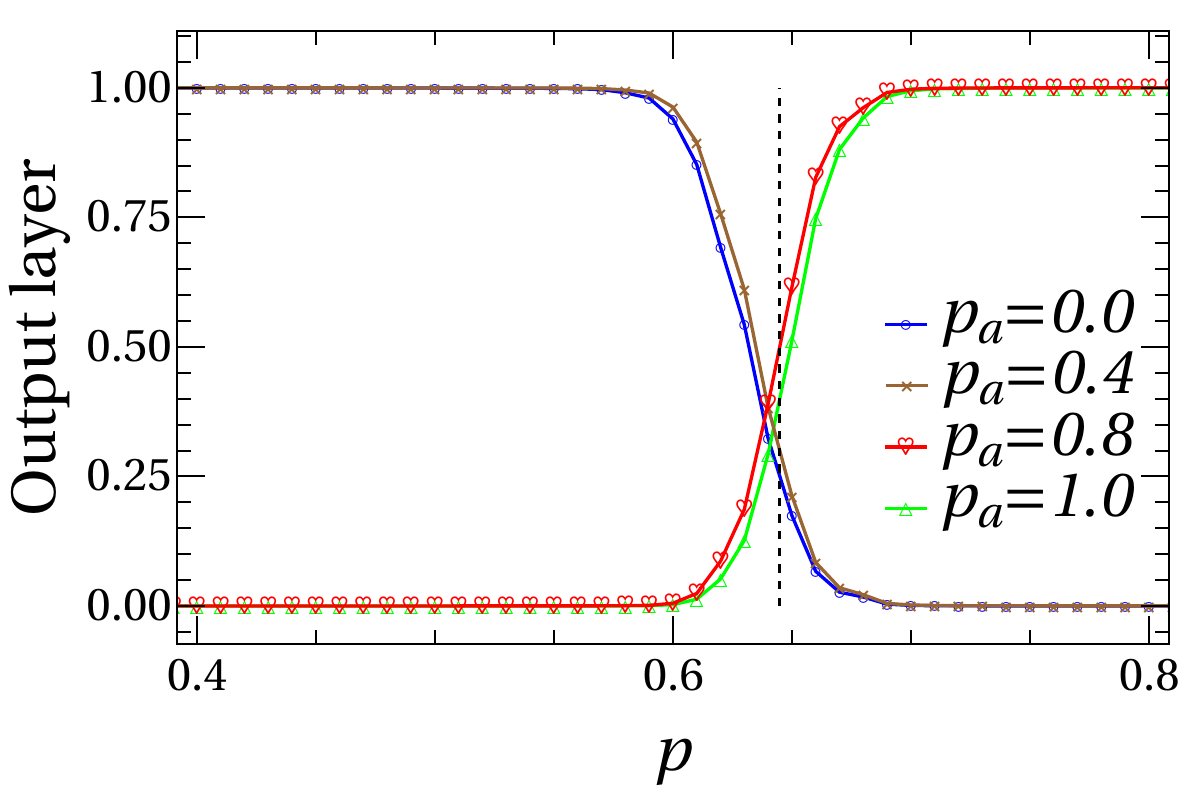}}
\subfigure[$L= 48$]{
\label{Fig.48.}
\includegraphics[height=0.29\textwidth]{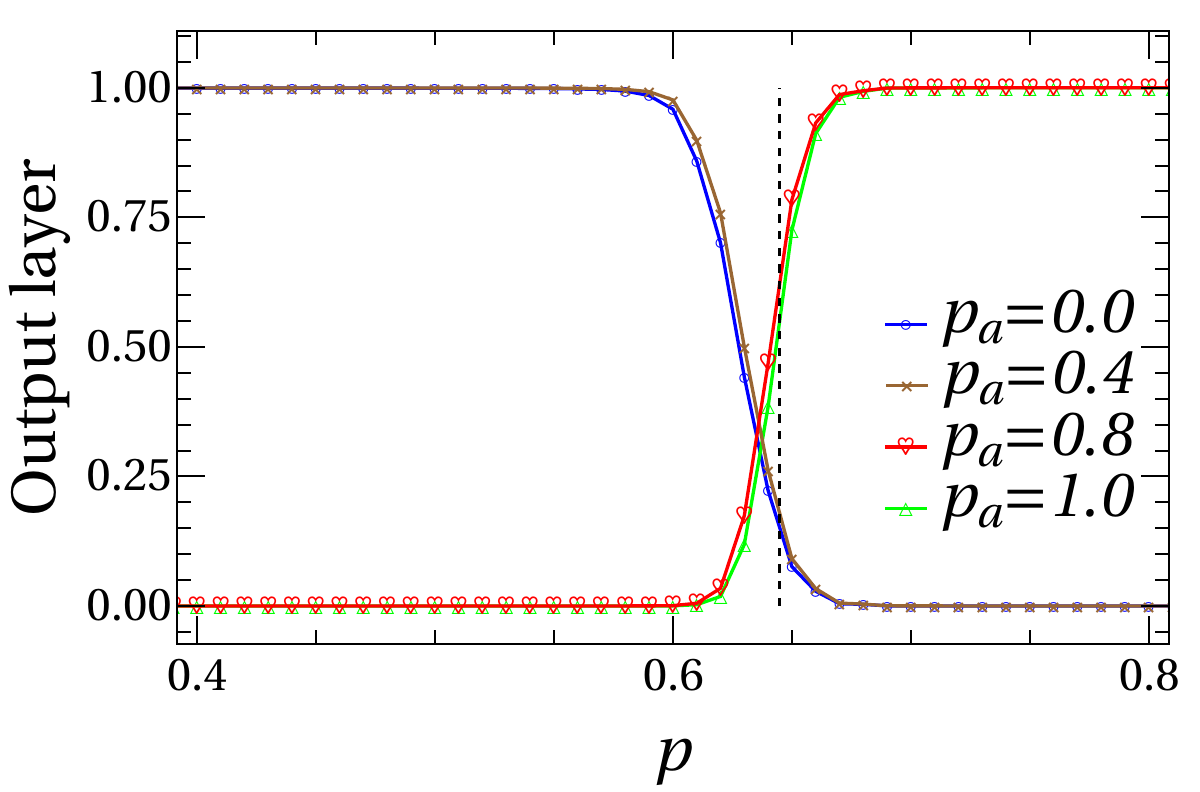}}
\subfigure[$L= 64$]{
\label{Fig.64.}
\includegraphics[height=0.29\textwidth]{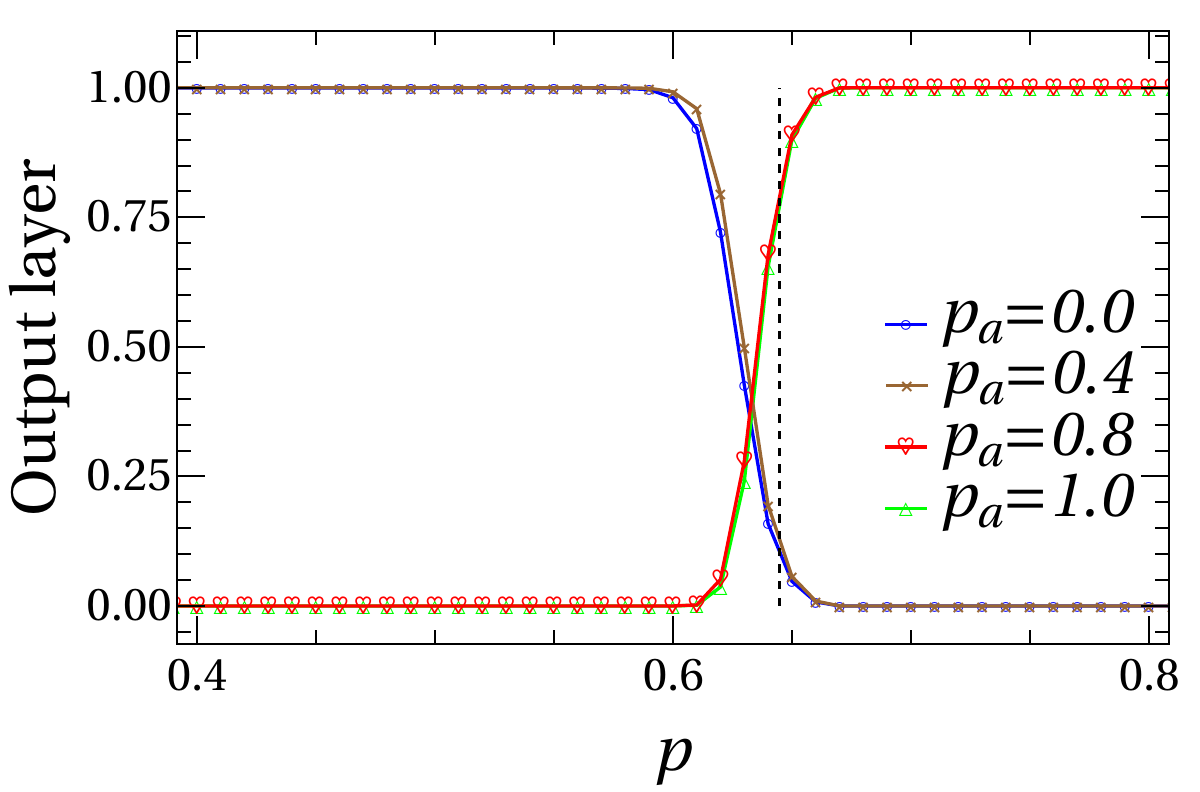}}

\caption{The output of SNN with the test anchor $p_{a} = 0,0.4,0.8$ and $1$, at different sizes (a) $L=16$, (b) $L=32$, (c) $L=48$ and (d) $L=64$.}
\label{fig123}
\end{figure}

Fig.~\ref{fig123} shows the similarity analysis of SNN with different anchors at $L= 16, 32,48$ and $64$. To study the dependence of anchors, we keep samples of $S_i$ from $p\in[0,1]$ but select anchor $p_{a} = 0,0.4,0.8$ and $1$ for $S_j$ for comparison. Since during training process the SNN learns nothing about the samples of $p_{a}=0.4$ and $0.8$, here it also contains the idea of transfer learning, i.e., the range $p\in[0,0.1]\cup [0.9,1]$ as "source domain" and $p_{a}=0.4$ or $0.8$ as "target domain". As shown in Fig.~\ref{fig123}, the similar curve $s$ changes with the different anchors.  For $p_{a} = 0$ (blue curve) and $0.4$ (brown curve), the similarity curve $s$ of all configurations is from "1" to "0" at $p\in[0,1]$, but for $p_{a} = 0.8$ (red curve) and $1$ (green curve), it becomes the opposite trend as from "0" to "1". Because the configurations of $p_{a} = 0$ and $0.4$ have similar phase to the ones of $p\in[0,0.1]$, but totally different at $p_{a} = 0.8$ and $1$.

\subsection{Extrapolated results of critical points}
\begin{figure}[htbp]
\centering
\subfigure[$p_{a} = 0$]{
\label{Fig.0.0}
\includegraphics[height=0.29\textwidth]{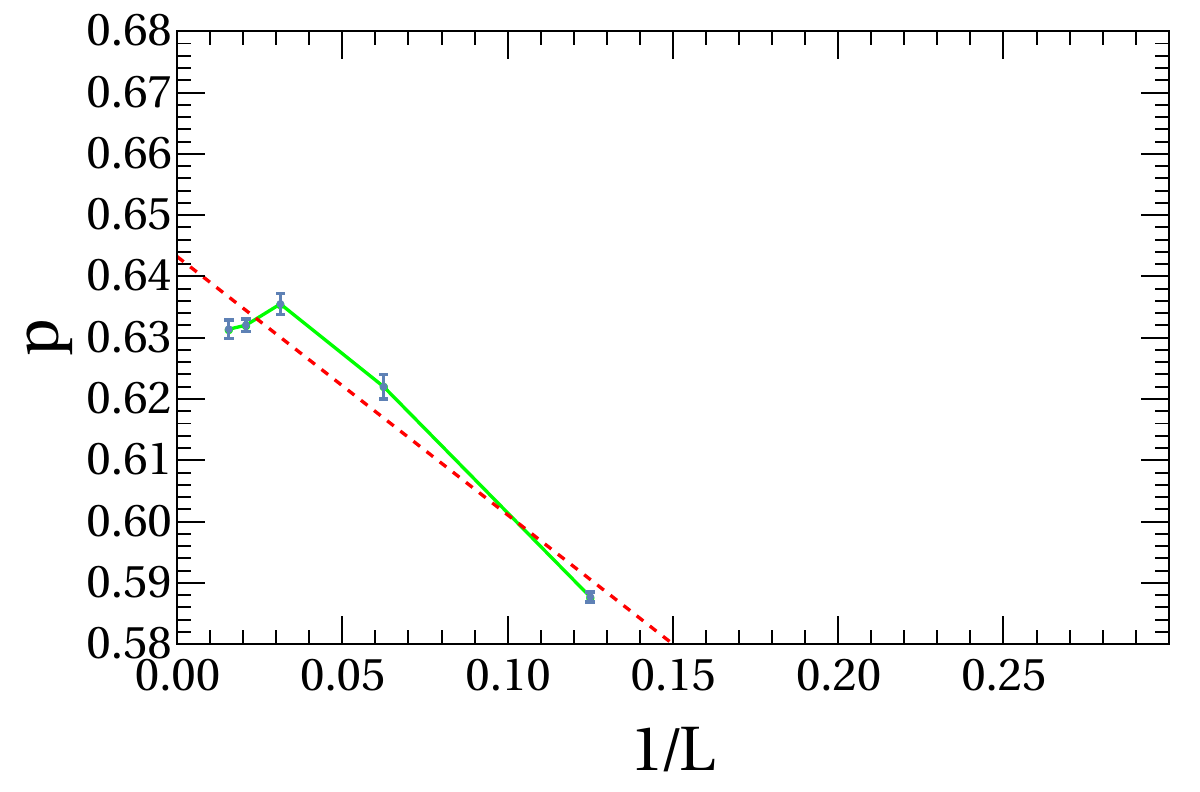}}
\subfigure[$p_{a} = 0.4$]{
\label{Fig.0.4}
\includegraphics[height=0.29\textwidth]{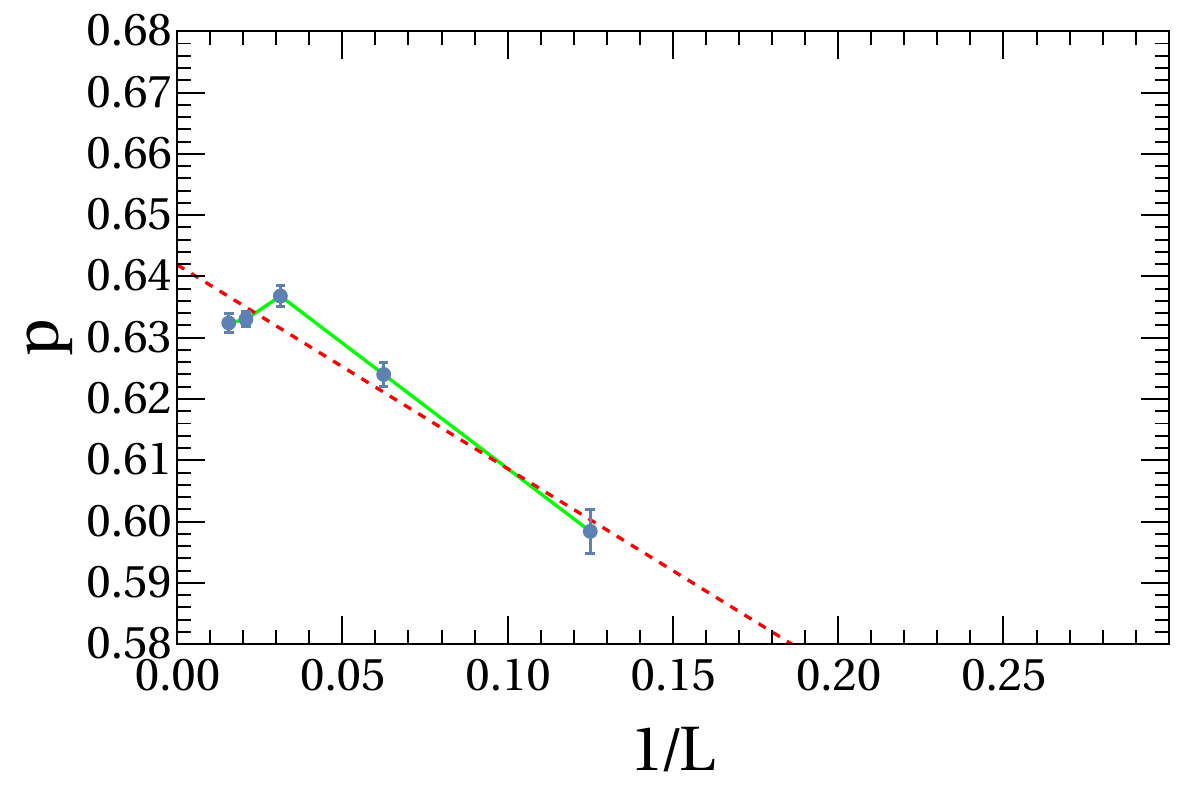}}
\subfigure[$p_{a} = 0.8$]{
\label{Fig.0.8}
\includegraphics[height=0.29\textwidth]{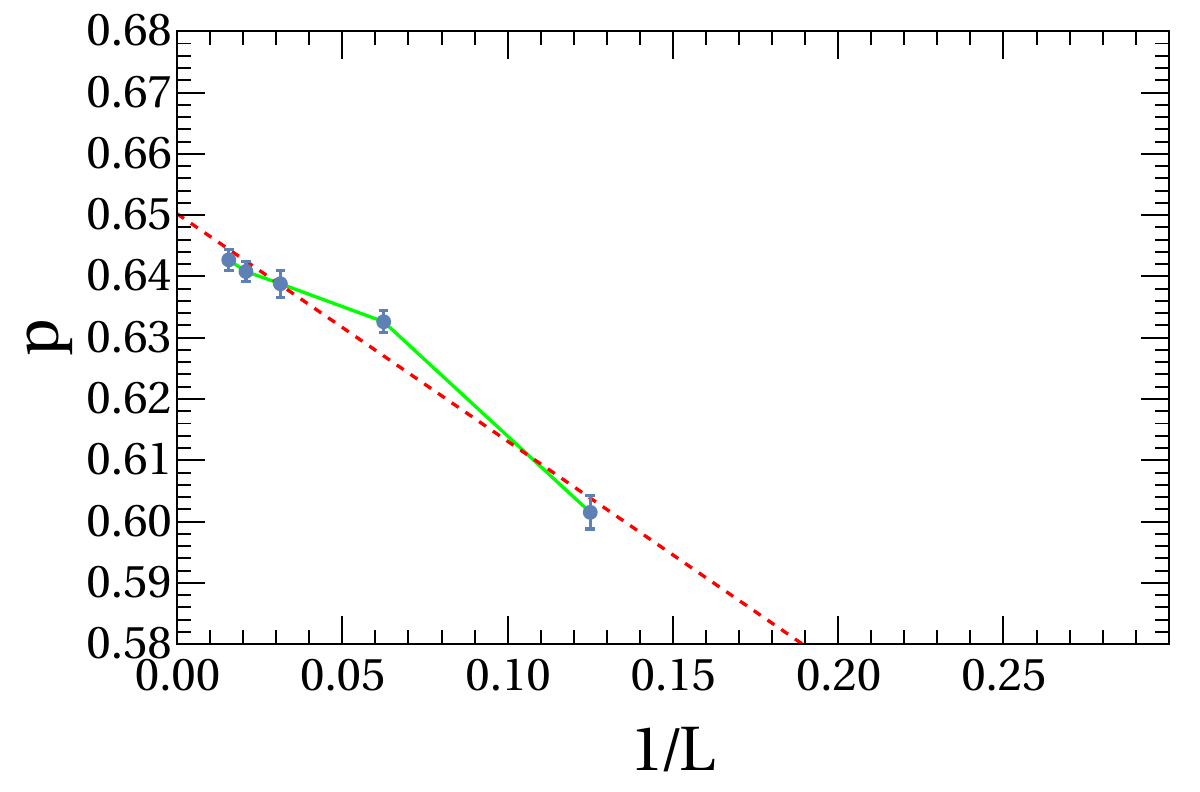}}
\subfigure[$p_{a} = 1$]{
\label{Fig.1.0}
\includegraphics[height=0.29\textwidth]{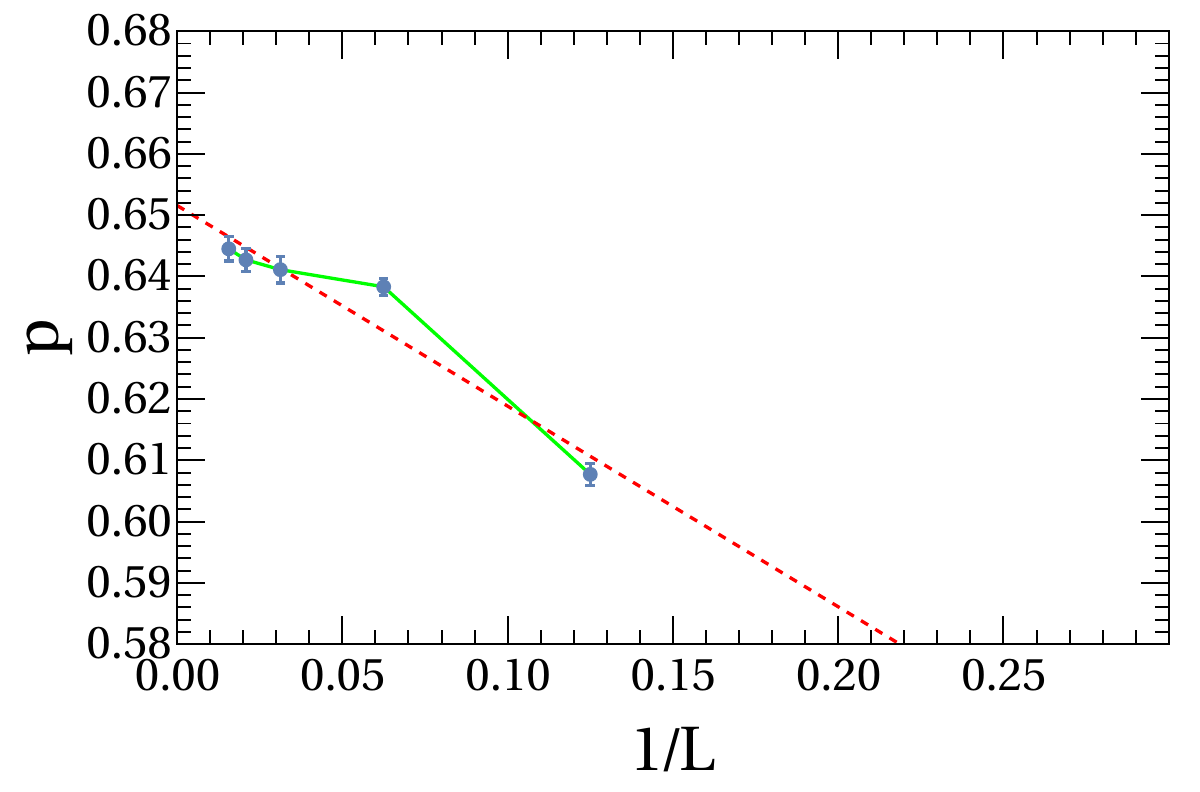}}

\caption{Extrapolation of the critical probability $p_c$ to infinite lattice size with anchor (a) $p_{a} = 0$, (b) $p_{a} = 0.4$, (c) $p_{a} = 0.8$ and (d) $p_{a} = 1$. The predictions of this $p_c^{\infty}$ are summarized in Table.~\ref{tab:initial}. The results for each system size are obtained by running the SNN five times independently. The confidence interval for $p_c^{\infty}$ is determined by the Weighted Least-Squares Estimation method.}
\label{figlimit}
\end{figure}

To deeply analyze the dependence of the critical point predicted by SNN on anchor points, 
we extrapolate the results at $L=8,16,32,48$ and $64$ to zero on the $1/L$ scale under anchor points $p_{a} = 0,0.4,0.8$ and $1$, by finite-size scale fitting (FSS)~\cite{binder1987finite,fisher1972scaling,privman1988finite,fan2007determination}. As shown in Fig.~\ref{figlimit},  the critical values increase with the size of $L$, which is consistent with the layer of phase transition models under limited scale. The SNN results of $p_c$ at infinite lattice size are summarized in Table.~\ref{tab:initial}. Comparing with the theoretical value $p_c=0.644700185(5)$~\cite{jensen1999low}, it shows the SNN can relatively predict critical value $p_c^{\infty}$ within deviation. 
The closest result of SNN is $p_c=0.6426$ at anchor $p_a=0.4$, and the SNN results  increase with the value of selected anchors, which also indicates the results of SNN may have anchor dependence. However, this achor dependence can be diluted by expanding the training set, which will be discussed in Subsection~\ref{extension}.

\begin{figure}[ht]
\centering
\subfigure[$p_a=0$]{
\label{Fig.1}
\includegraphics[height=0.29\textwidth]{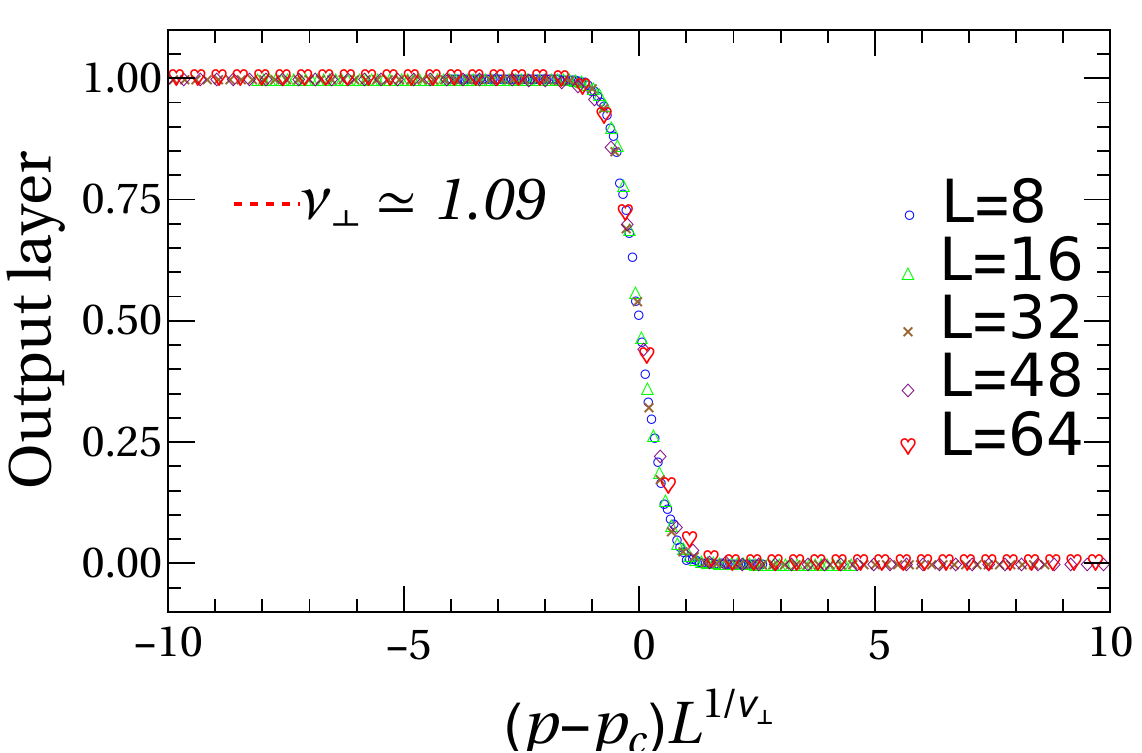}}
\subfigure[$p_a=0.4$]{
\label{Fig.2}
\includegraphics[height=0.29\textwidth]{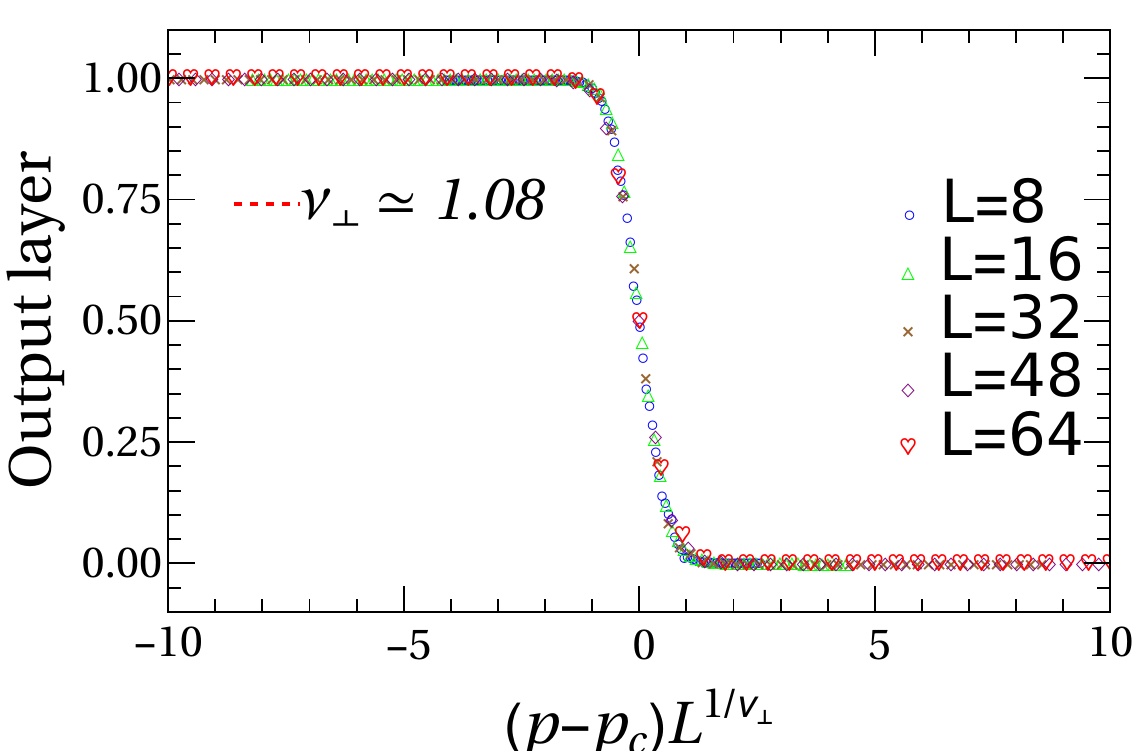}}
\subfigure[$p_a=0.8$]{
\label{Fig.3}
\includegraphics[height=0.29\textwidth]{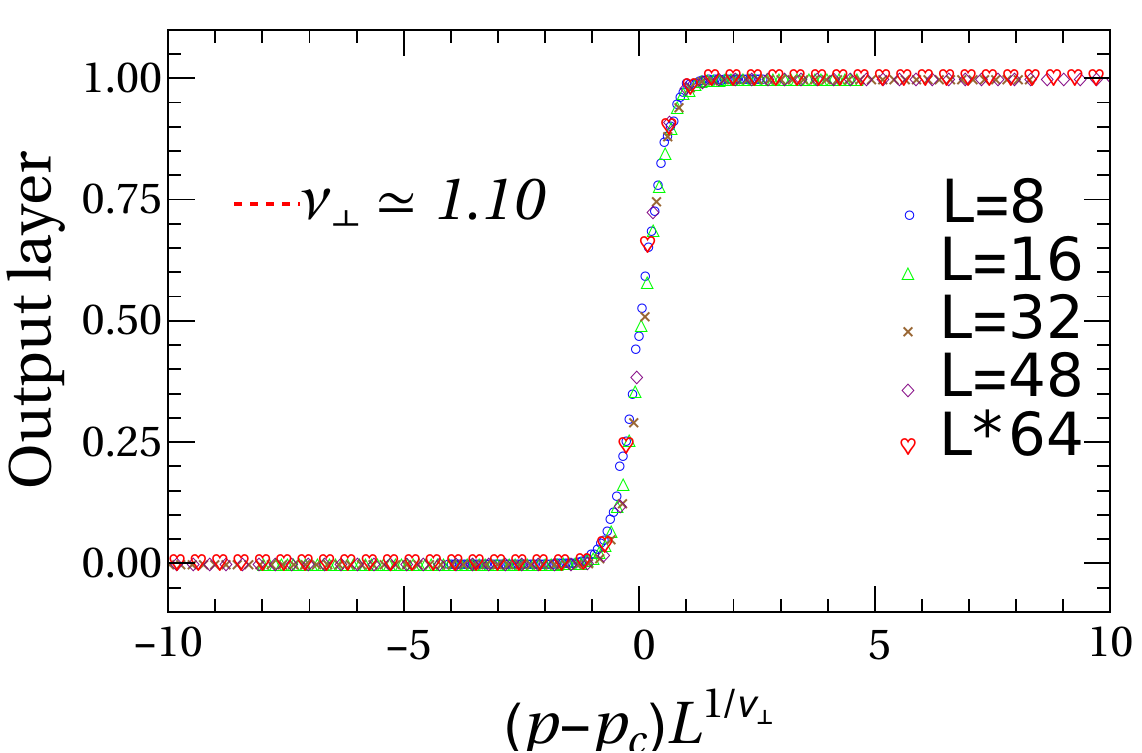}}
\subfigure[$p_a=1$]{
\label{Fig.4}
\includegraphics[height=0.29\textwidth]{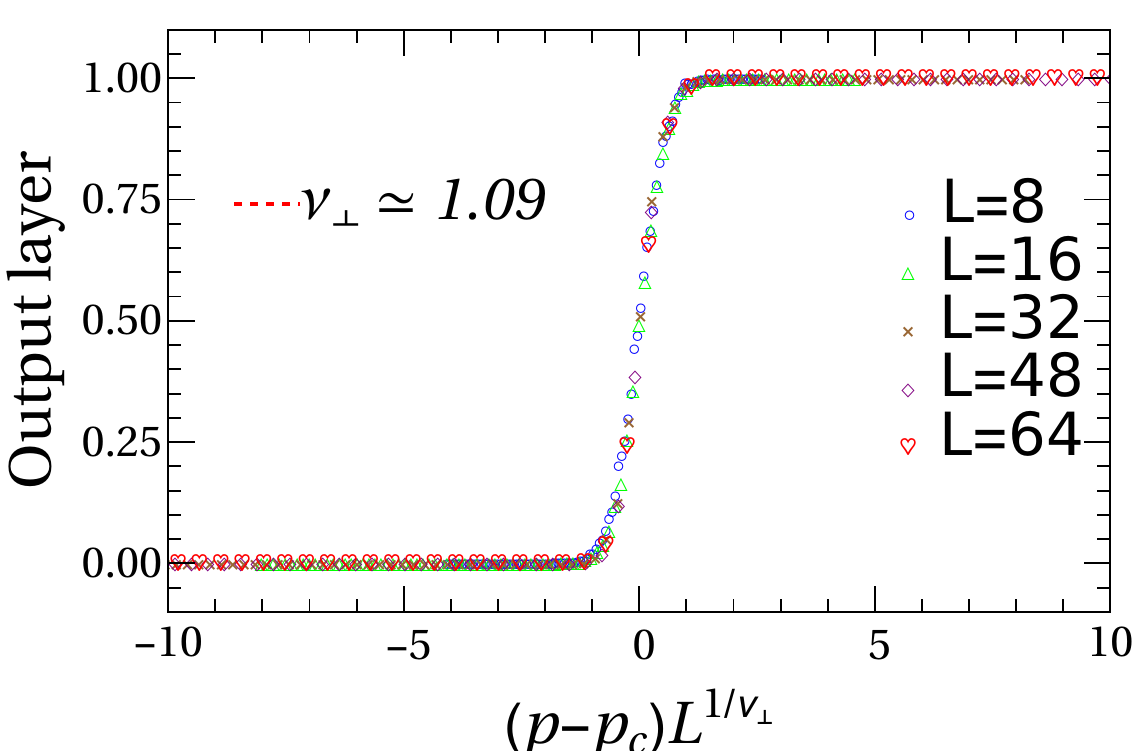}}
\caption{The collapse of the average output layer as a function of $(p- p_{c}) L^{1/ \nu_{\perp}}$, with anchor (a) $p_{a} = 0$, (b) $p_{a} = 0.4$, (c) $p_{a} = 0.8$ and (d) $p_{a} = 1$.}
\label{fignu}
\end{figure}

Further, the curves of $P_0$ in Figure.~\ref{fig123} can be fitted with a sigmoid function:
\begin{equation}
 p \to \frac1{1+e^{-\frac{(p-p_c)}{\sigma}}} \quad \text{or} \quad  p \to 1- \frac1{1+e^{-\frac{(p-p_c)}{\sigma}}},
  \label{eq:sigmoid_DP}
\end{equation}
for $1-s$ or $s$, where $\sigma$ is the scaling width. As shown in Fig.~\ref{fignu}, by the technique of data collapse we can obtain the spatial correlation exponent $\nu_\perp$ via the scaling $(p-p_c) L^{1/\nu_\perp}$. From $P_0$ at lattice sizes of $L =8, 16, 32, 48$ and $64$, a proper value of $\nu_\perp$ can be found to fit the scaling. Table.~\ref{tab:initial} shows the fitted $\nu_\perp$ of each fixed anchors, all of which are consistent with the theoretical value $1.09$~\cite{doi:10.1080/00018730050198152}. Since $\nu_\perp$ has more dependence on raw configuration and also has lower precision, it is not as sensitive to changes with the anchors as $p_c^{\infty}$.

\begin{table}[htbp]
\caption{The critical value $p_c^{\infty}$ and spatial correlation exponent $\nu_{\perp}$ estimated by SNN with archor $p_a=0,0.4,0.8$ and $1$.}
\label{tab:initial}
\centering
\resizebox{0.6\linewidth}{!}{
\begin{tabular}{lcccccccc}
\hline\hline
         &$p_a=$0.0 &$p_a=$0.4   &$p_a=$0.8       &$p_a=$1.0    \\ 
\hline
\quad $p_c^{\infty}$  &$0.6388 \pm 0.0003 $& $0.6420 \pm 0.0013$  &$0.6477 \pm 0.0011$  &$0.6505 \pm 0.0013$        \\
\quad$\nu_{\perp}$    &1.09 & 1.08  &1.1  &1.09      \\
\hline\hline
\end{tabular}}
\end{table}

\subsection{The SNN results of Ising model}
\begin{figure}[ht]
\centering
\subfigure[Different $p_a$]{
\label{Fig.11}
\includegraphics[height=0.29\textwidth]{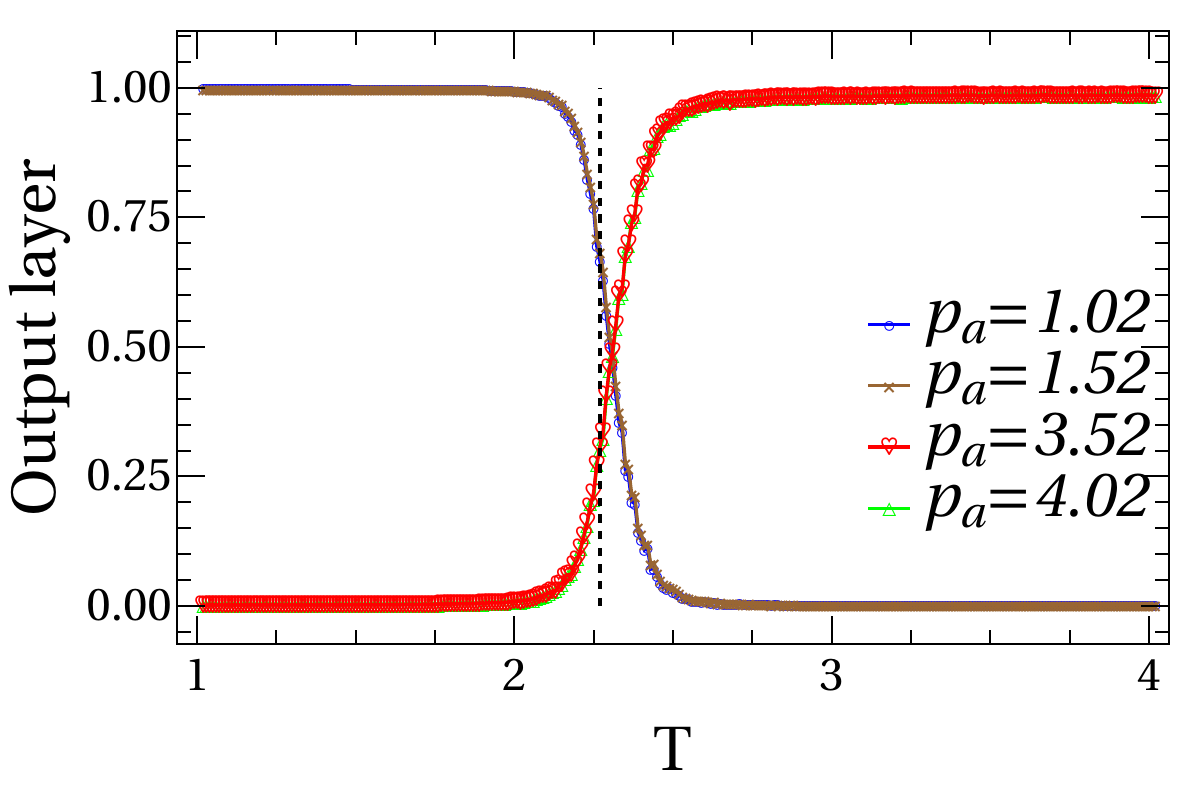}}
\subfigure[$p_a=1.02$]{
\label{Fig.22}
\includegraphics[height=0.29\textwidth]{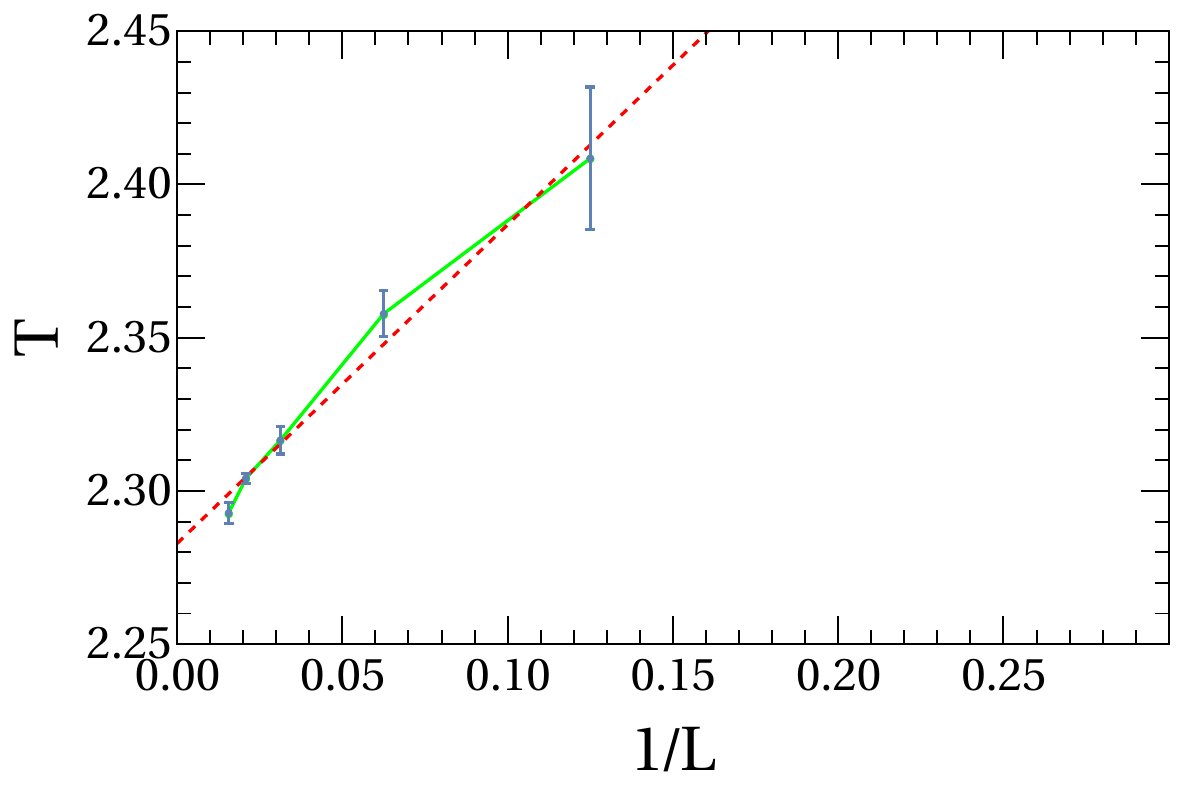}}
\subfigure[$p_a=1.02$]{
\label{Fig.33}
\includegraphics[height=0.29\textwidth]{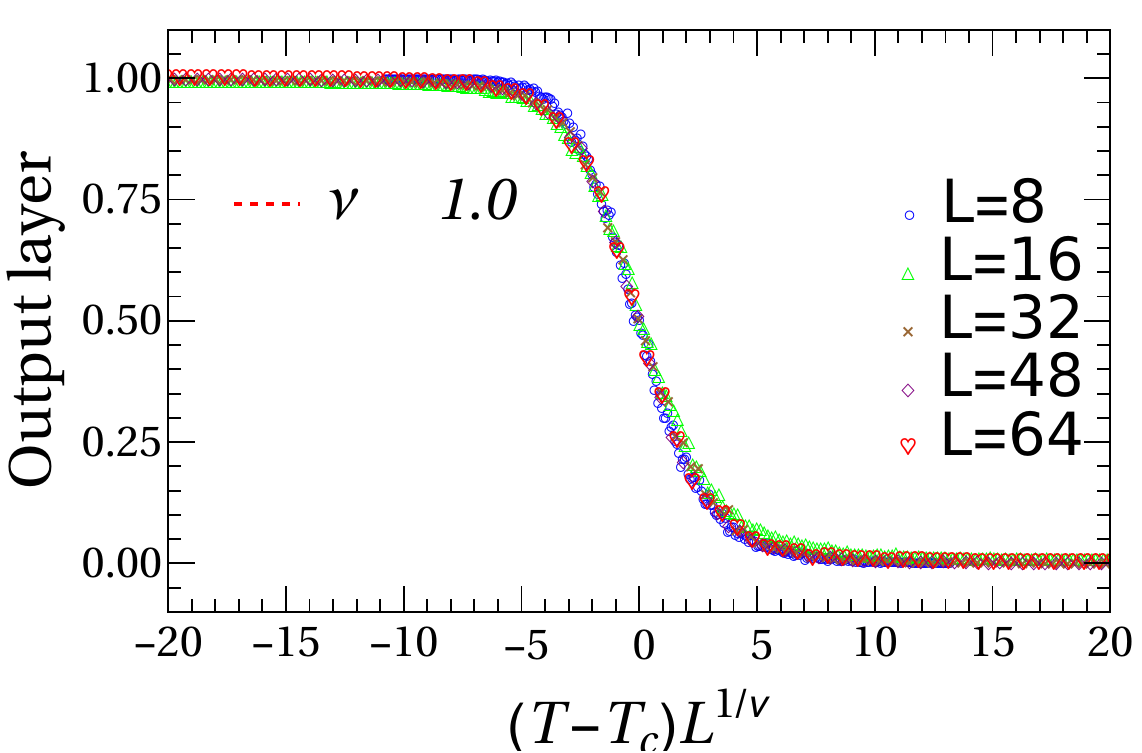}}
\subfigure[$p_a=2.269$]{
\label{Fig.44}\includegraphics[height=0.29\textwidth]{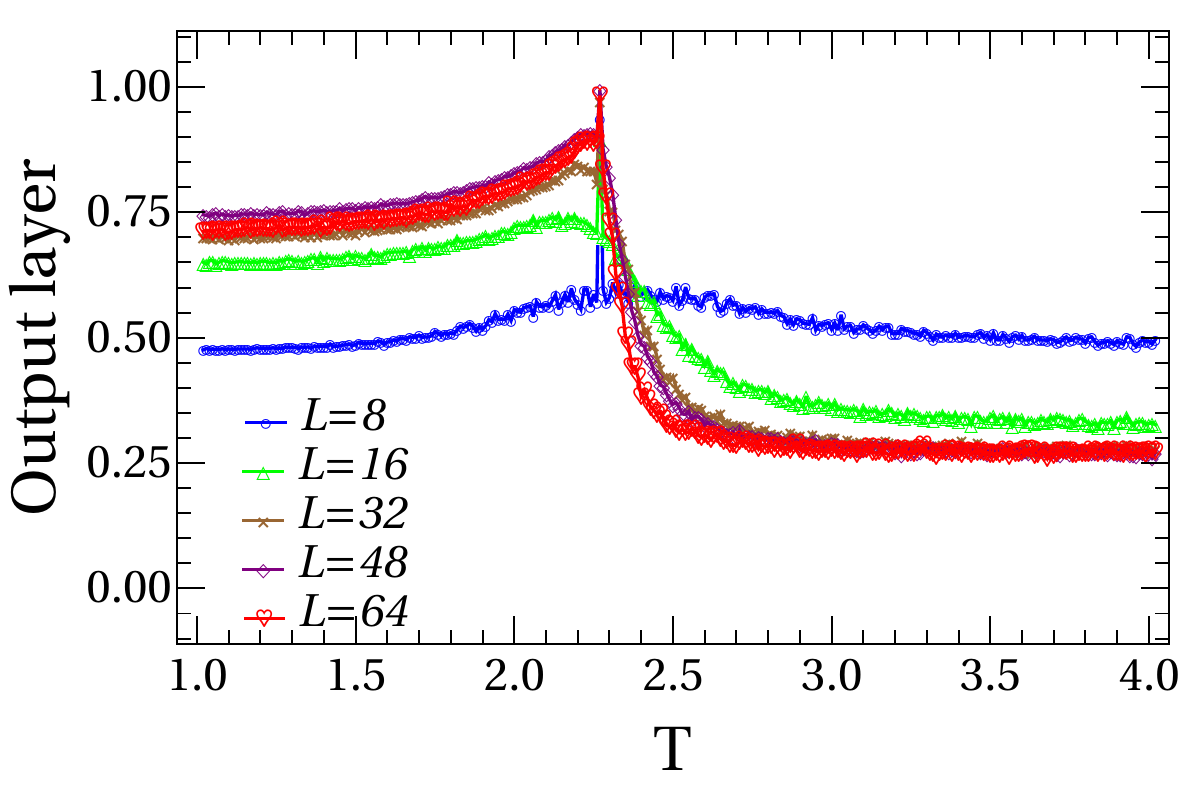}}
\caption{The SNN results of Ising model. (a) The output of SNN with the test anchor $p_{a} = 1.02$ when $L = 32$, (b) Extrapolation of the critical probability $T_c$ to inﬁnite lattice size with anchor (a) $p_{a} = 1.02$, (c) The collapse of the average output layer as a function of $(T- T_{c}) L^{1/ \nu_{\perp}}$, with anchor $p_{a} = 1.02$, (d) The output of SNN with anchor $p_a=2.269$ at $L = 8, 16, 32, 48$ and $64$.}
\label{ising}
\end{figure}

To evaluate the cross-system robustness of the SNN method, we apply it to the two-dimensional Ising model. Fig.~\ref{ising} presents the results of the Ising model analysis using the SNN. 
Similar to the SNN learning approach for DP, we choose different anchor points of temperature. To simplify the analysis, we select anchor points as $p_a = 1.02,1.52,3.52$ and $4.02$, which are the boundary values of the entire labeled temperature range $T\in[1.02,1.52]\cup [3.52,4.02]$.

In Fig.~\ref{ising} (a) shows the results of SNN for the four chosen anchor points at $L = 32$. Obviously, according to the studies in the previous subsections, these outputs are in line with expectations: the similarity $s$ increases to $1$ as the two compared configurations are from closer temperature values and it decreases to $0$ on the contrary.
The black dash line in Fig.~\ref{Fig.11} represents the theoretical critical value of the Ising model at $2.269$~\cite{onsager1944crystal}. Due to finite-size effects, it is a discrepancy between the intersection of the curves and the theoretical value.

To obtain the predicted critical value $T_c$ from the SNN, we perform the finite-size scaling analysis for the anchor point at 1.02 with system sizes of $L = 8, 16, 32, 48$, and $64$, as we have used for the (1+1) DP before. The results are shown in Fig.~\ref{Fig.22}, and the predicted critical temperature of infinite system is $T_c ^{\infty}= 2.2777 \pm 0.0033$, close to the theoretical one. And also, by the data collapse method, we achieve a good fitting of the similarity curves for the five system sizes as illustrated in Fig.~\ref{Fig.33}, and then obtain the correlation exponent $\nu = 1.0$ as the numerical solution in Ref.~\cite{cardy1996scaling}. In Fig.~\ref{Fig.44}, we present the SNN results by choosing a specific anchor point $p_a = 2.269$, close to the theoretical critical temperature. The similarity output exhibits a peak at the anchor point, indicating that at this temperature the input configurations are only similar to themselves. It is an interesting behavior and we will give detailed discussions in Section~\ref{anch_choice}.

\section{Discussion}\label{discussion}

\subsection{In-depth study on the selection of anchor points}
\label{anch_choice}

\begin{figure}[ht]
\centering
\begin{tabular}{ccc}   
    \includegraphics[width=0.30\textwidth]{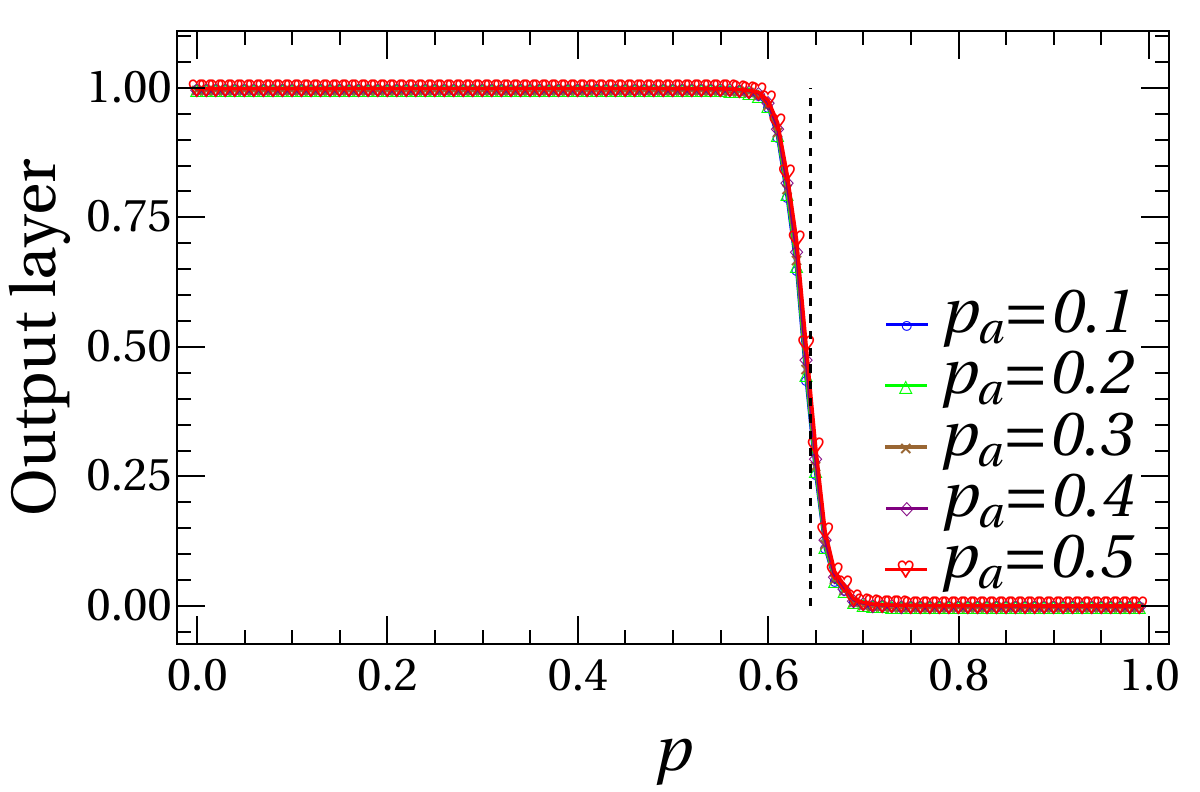} &
    \includegraphics[width=0.30\textwidth]{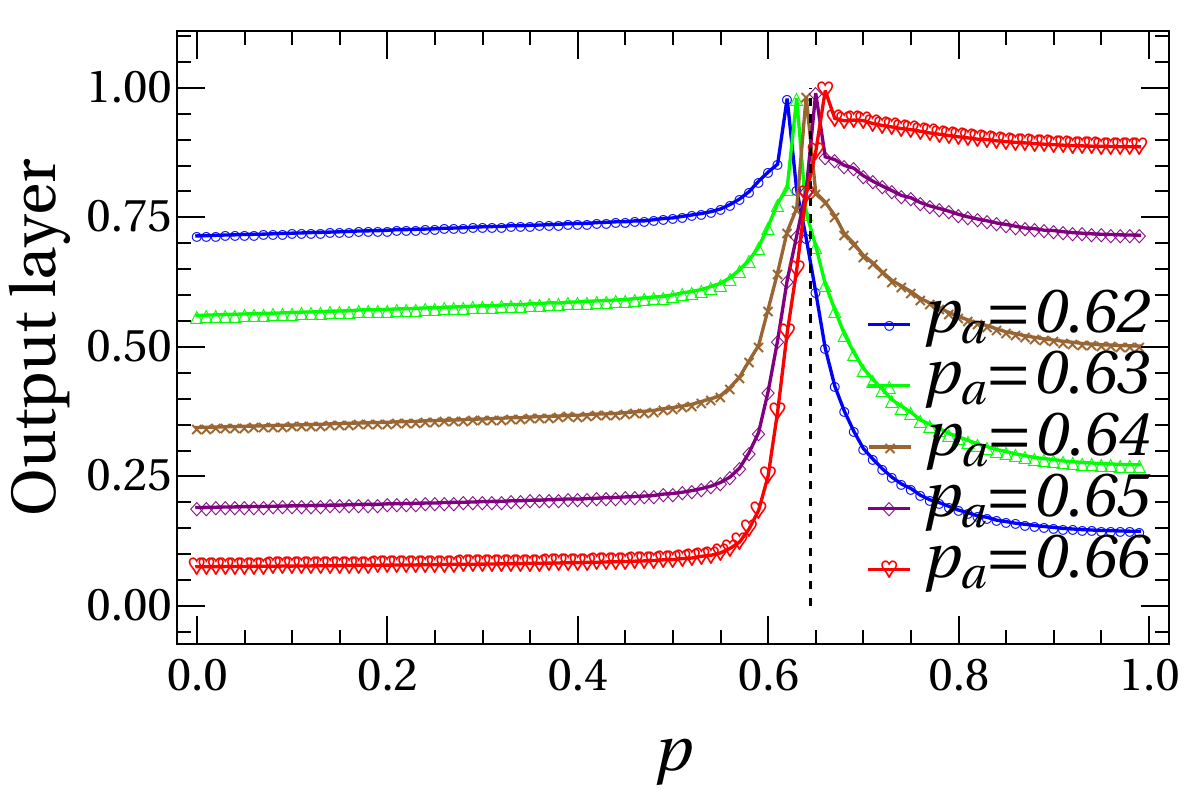}&
    \includegraphics[width=0.30\textwidth]{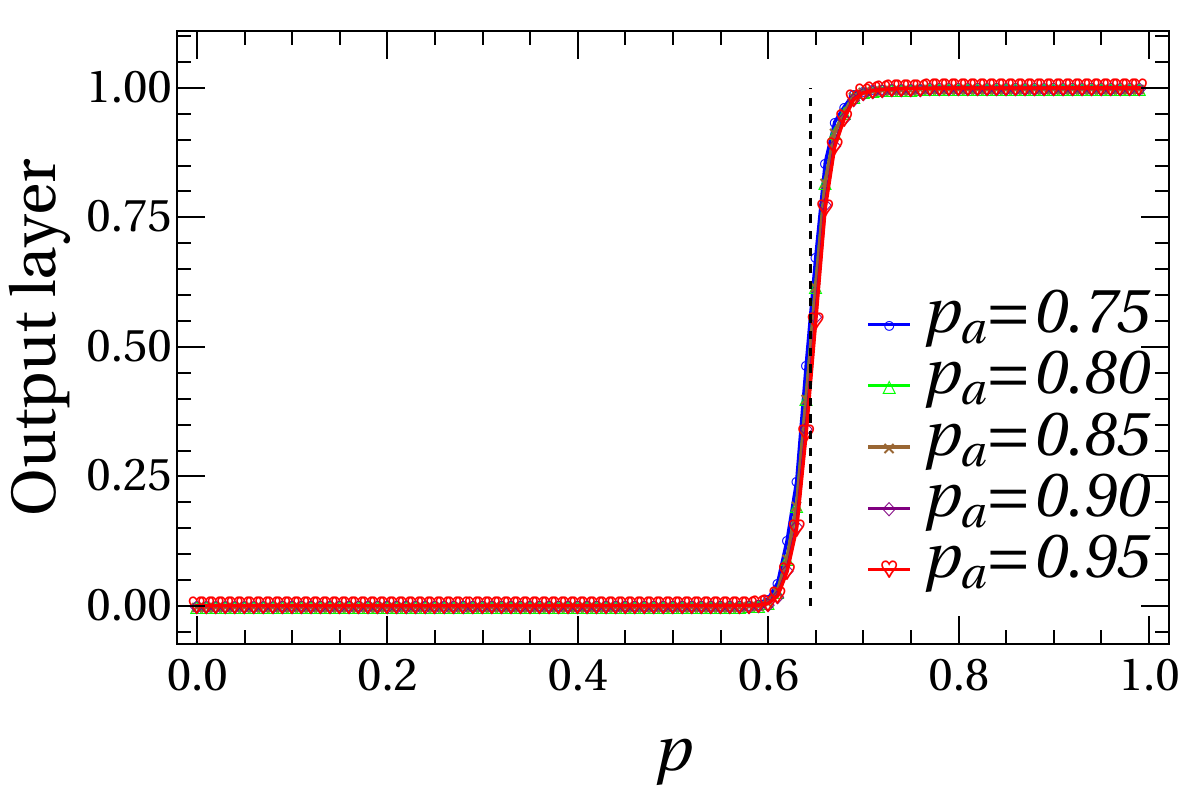}\\
     (a) &  (b) &  (c)
\end{tabular} 
\caption{The SNN results of the (1+1) dimensional directed percolation model by choosing different anchor points at $L = 32$. (a) $p_{a} < p_{c}$, (b) $p_{a}$ approaches $p_{c}$ , (c) $p_{a} > p_{c}$. The black dashed line represents the theoretical critical value $p_c = 0.644700185(5)$.}
\label{choice}
\end{figure}

To deeply discover the relationship between the predicted critical point and the choice of anchor points, here we conducted more tests using different anchor points, taking the (1+1) dimensional directed percolation model as an example. In section~\ref{anchor_dependence}, we have selected four points deviating from the critical point as anchor points $p_a=0, 0.4, 0.8,$ and $1$ to determine the system's critical value and spatial correlation exponent through finite-size scaling and data collapse. Now to observe the impact of anchor points on critical value prediction over a broader range, we supplement with additional anchor points $p_a=0.1, 0.2, 0.3, 0.4,$ and $0.5$, as shown in Fig.~\ref{choice}. For the results of SNN, it can be seen that the similarity $s$ decreases progressively from $p=0$ to $1$, and the outputs of five curves are nearly consistent. From Fig.~ \ref{choice}, it is evident that the critical points predicted by the SNN closely match the theoretical value $p_c = 0.644700185(5)$~\cite{jensen1999low}. As in Section
~\ref{anchor_dependence}, by the skill of finite-size scaling and data collapse, we also can obtain the critical value and spatial correlation exponent of infinite system. In Fig.~\ref{choice}(c), by using $p_a = 0.75,0.8,0.85,0.9$ and 0.95, the SNN also can gives the critical information of system, similar to Fig.~\ref{choice}(a).

In contrast, Fig.~\ref{choice}(b) displays entirely different behavior compared to Fig.~\ref{choice}(a) and \ref{choice}(c). In Fig.~\ref{choice}(b), we chose anchor points at $p_a=0.62, 0.63, 0.64, 0.65,$ and $0.66$, in the critical region. In these cases, the curves $s$ are not sigmoid-like function but have a structure of single peak. The maximum values
of the peak for each lattice size $L$ are all close to 1. The reason of this behavior is that the configurations close to
the critical point are only very similar to themselves but different from the ones of other bond probabilities. As the
value of $p$ moves away from the anchor $p_a =  0.62, 0.63, 0.64, 0.65$ and $0.66$, the similarity curves decrease rapidly within a short distance in the
range of critical area and then changes slowly.
Because of fluctuations near the critical region in the system, the SNN results also show a degree of deviation from the  perspective of $x$-coordinates $p$. 

The above analysis demonstrates that it is very challenging to obtain critical information of the system by using points in the critical region as anchor points. Hence, in the main text, we opted for points far from the critical region as anchor points.

\subsection{The optimal training set}\label{extension}
In previous tests of the (1+1) dimensional directed percolation model, we sampled configurations $S_i$ and $S_j$ in the range of $p\in[0,0.1]\cup [0.9,1]$ as the training set of SNN, to minimize human intervention. However, from Section~\ref{sec:Results} it is obviously that the expected critical region is very narrow, which makes it possible to expand the training set. Typically, a wider and more accurate label training set is beneficial to improving the accuracy of NN predictions. Inspired by Ref.~\cite{shen2022transfer}, an iterative approach is introduced to find an optimal interval of the training set. By defining the initial training set as $[0, l^{(0)}]\cup[r^{(0)}, 1]$ with $l^{(0)}=0.1$ and $r^{(0)}=0.9$, the updating process can be started with the initial estimate critical probability $p_{c}^{(0)}$:
\begin{equation}
l^{(i+1)}=\frac{l^{(i)}+p_{c}^{(i)}}{2},\quad 
r^{(i+1)}=\frac{r^{(i)}+p_{c}^{(i)}}{2},
\label{equ.lr}
\end{equation}
where $i$ represents the $i$-th expansion. $p_{c}^{(i)}$ is the estimate critical probability by SNN on the $i$-th expansion. For each iteration, it has to check whether at least 99\% of the output similarity in the $i$-th interval of training is classified as category ``0'' or ``1''. If the condition is met, it would continue the process, otherwise make a correction like $l^{(i+1)}\to l^{(i+1)}/2$. The iteration would be stopped when the interval of training set can't be extended any more.

\begin{figure}[htbp]
\centering
\includegraphics[width=0.6\textwidth]{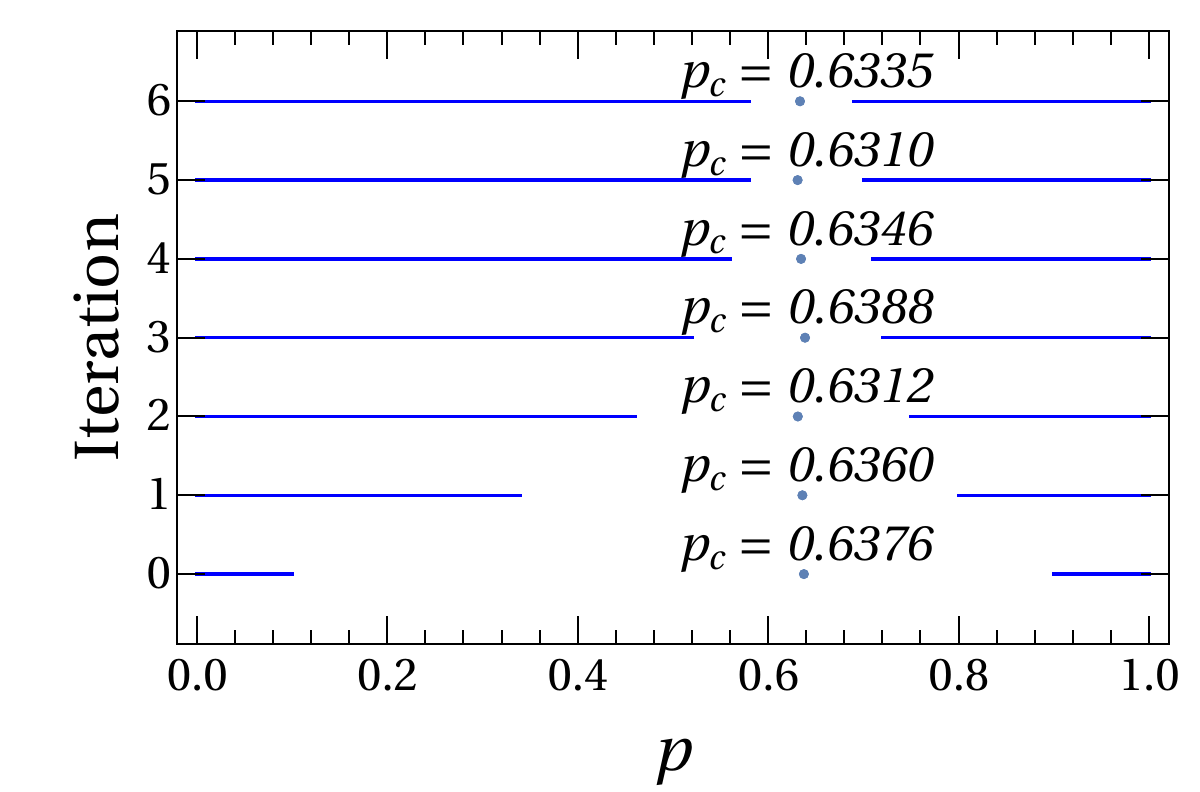}
\caption{Evolution of the training set, and the corresponding critical bond probability values with DP system size $L=32$}.
\label{fig:iteration}
\end{figure}

\begin{figure}[tbp]
\centering
\subfigure[]{
\label{pr0.0_L_diffsize_limit_iteration}
\includegraphics[height=0.29\textwidth]{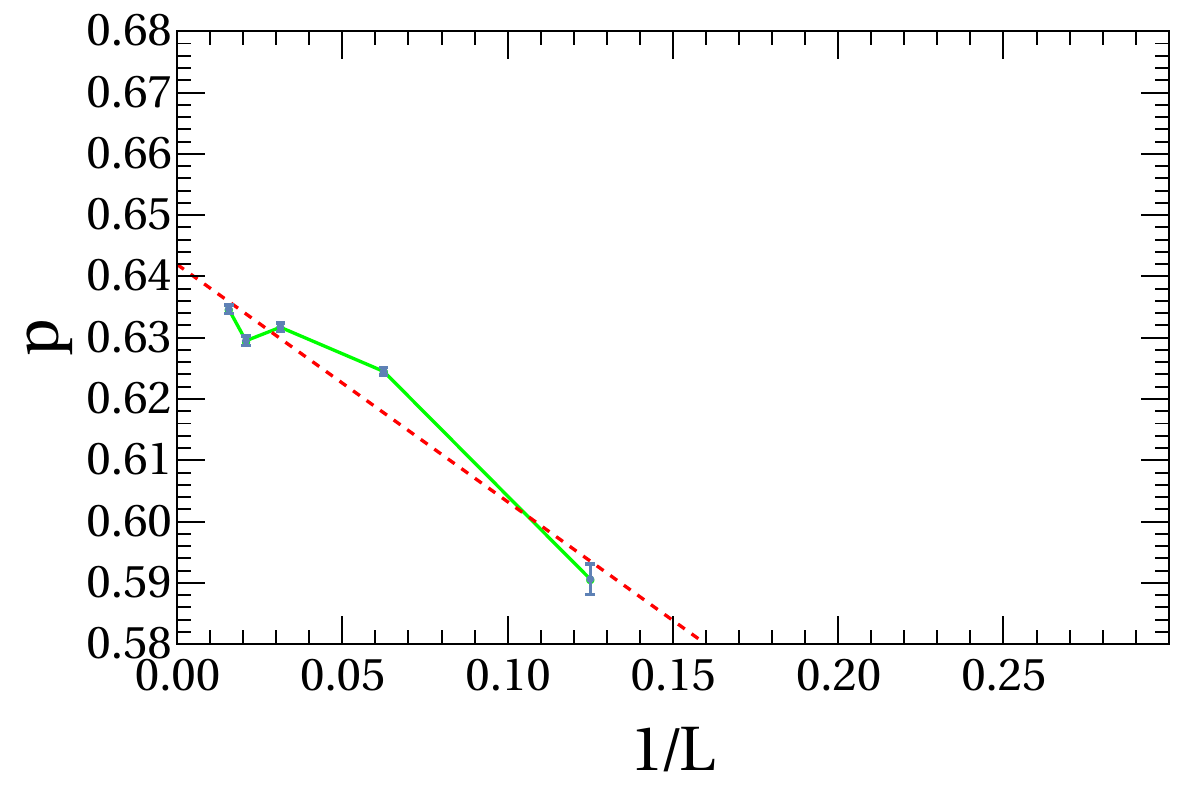}}
\subfigure[]{
\label{Datacollapsepc_pr0.0_v_iteration}
\includegraphics[height=0.29\textwidth]{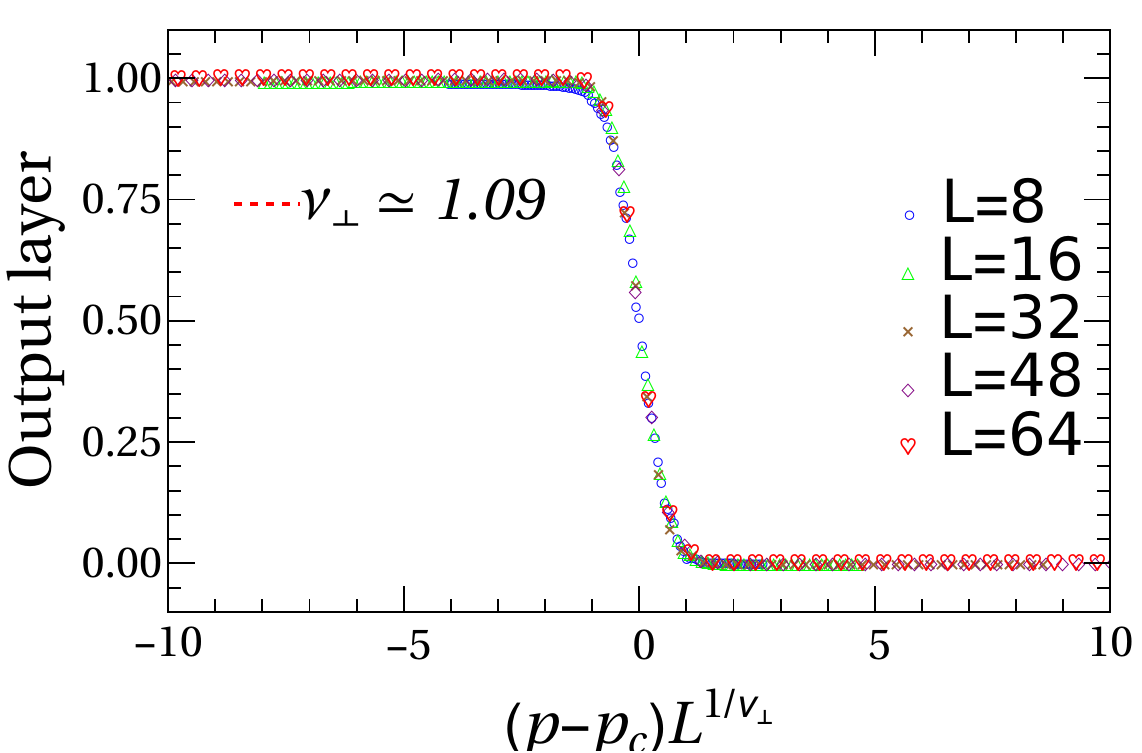}}
\caption{(a) Extrapolation of the critical probability $p_c$ to inﬁnite lattice size with $p_{a} = 0$, by the optimal training intervals in Table~\ref{tab:optimal}. (b) The collapse of the average output layer as a function of $(p- p_{c}) L^{1/ \nu_{\perp}}$.}
\label{extension_iteration}
\end{figure}

In Fig. \ref{fig:iteration}, we show the iteration process to achieve the estimate critical value $p_{c} = 0.6335$ of optimal training set $[0,0.58]\cup[0.69,1]$ for $L = 32$, with anchor $p_a = 0$. Obviously, the optimal training interval is much larger than the initial one, and the final result would also be more convinced. Table~\ref{tab:extension} gives the number of iterations, the optimal training interval and the critical probability predicted by SNN of $p_a = 0$ for lattice size $L=8, 16, 32, 48$ and $64$. Then, the critical probability of infinite size can also be obtained by the extrapolation in Fig.~\ref{pr0.0_L_diffsize_limit_iteration}, as $p^{\infty}_c=0.6420$, and also by the data collapse we can get $\nu_\perp\simeq 1.09$ in Fig.~\ref{Datacollapsepc_pr0.0_v_iteration}. Table~\ref{tab:optimal} shows the critical probabilities predicted by the optimal training interval under anchor point $p_a=0,0.4,0.8$ and $1$. Compared to the results of initial training interval in Table~\ref{tab:initial}, a larger training set makes the SNN learn features more accurate, from which the results are less affected by the selected anchor values. Based on the previous summary, it can be inferred that by using the optimal training interval, selecting any anchor in training interval as test sample set can predict a reliable result. On the other hand, Table~\ref{tab:optimal} also lists the results of supervised learning~\cite{PhysRevE.103.052140}, autoencoder (AE)~\cite{PhysRevE.103.052140} and DANN~\cite{shen2022transfer} for comparison.

\begin{table}[htbp]
\caption{The optimal training interval of $L=8, 16, 32, 48$ and $64$, with iteration step and predicted $p_c$ at each size.}
\label{tab:extension}
\centering
\resizebox{0.7\linewidth}{!}{
\begin{tabular}{lcccc}
\hline\hline
{Lattice size} &Iteration step&Optimal training interval  &$p_{c}$ \\ 
\hline
L = 8  &2  &$[0, 0.18]\cup[0.82,1]$ &0.6006  \\
\hline
L = 16 &6  &$[0, 0.53]\cup[0.73,1]$ &0.6258  \\
\hline
L = 32 &6  &$[0, 0.58]\cup[0.69,1]$ &0.6335  \\
\hline
L = 48 &6  &$[0, 0.59]\cup[0.67,1]$ &0.6323  \\
\hline
L = 64 &6  &$[0, 0.60]\cup[0.67,1]$ &0.6357  \\
\hline\hline
\end{tabular}}
\end{table}

\begin{table}[htbp]
\caption{Comparing the supervised machine learning, autoencoder (AE), domain adversarial neural network (DANN), and SNN methods for calculating the critical value $p_{c}^{\infty}$ and spatial correlation exponent $\nu_{\perp}$ with optimal training intervals.}
\label{tab:optimal}
\centering
\resizebox{1\linewidth}{!}{
\begin{tabular}{lcccccccc}
\hline\hline
Method        &Supervised~\cite{PhysRevE.103.052140} &AE~\cite{PhysRevE.103.052140} &DANN~\cite{shen2022transfer}       &$p_a=0.0$ &$p_a=0.4$   &$p_a=0.8$       &$p_a=1.0$    \\ 
\hline
\quad$p_{c}^{\infty}$       &0.6408     & 0.643(2)     &0.6453(5)  &0.6420 & 0.6429  &0.6452  &0.6440         \\
\quad$\nu_{\perp}$ &1.09(2)    & None         &1.09(6)    &1.09 & 1.08  &1.1  &1.09      \\
\hline\hline
\end{tabular}}
\end{table}

\section{Conclusion}\label{conclusion}

In this paper, we have applied a Siamese Neural Network (SNN) to investigate the critical behavior of two prominent phase transition models: the (1+1) dimensional bond directed percolation of non-equilibrium system, and the 2-dimensional Ising model as a classic example of equilibrium phase transition. Here the SNN is used as a semi-supervised learning method, in which the key information of critical behavior can be predicted by labeling only a part of raw configurations as input for training. And the output of the SNN is similarity, which transforms the classification problems of phase transition into a discussion of difference, broadening the application of Machine learning (ML) techniques in the study of critical behaviors.

By the output similarity of SNN, we predict the critical points of the system at different lattice sizes, and then use the fitting based on finite-size scaling to to obtain the theoretical values at unlimited system. Further, the curve of similarity also can be fit to a sigmoid-like function, which allows us to calculate the spatial correlation exponent $\nu_{\perp}$ of the (1+1) dimensional bond directed percolation and correlation exponent $\nu$ of the 2-
dimensional Ising model in phase transition systems through data collapse.

During the test, the selection of anchor points is of particular importance. By selecting the same training set, we discussed the impact of configurations under different anchors on the results. As the chosen anchor is outside the critical area, the predicted critical point increases with the value of anchor. For the anchor close to critical point, the curve of similarity shows a structure of single peak at the anchor, which indicates these configurations at the anchor only $100\%$ similar to themselves. In this case, the critical point cannot be predicted by the SNN. To make the calculation more efficient and reduce anchor dependence, we optimize the training interval through an iterative method. After optimization, the critical point result obtained by SNN is very accurate, comparable to the results of traditional supervised learning, AE and DANN.

With the widespread utilization of machine learning techniques in the fields of statistics and condensed matter physics, it is a promising future for further advancements. The SNN approach introduced in this research paper has the potential to offer novel insights into the exploration of DP and Ising-like phase transitions.

\section*{Acknowledgements} 
This work was supported in part by Yunnan Fundamental Research Projects (Grant 202401AU070035), the Baoshan University Doctoral Research Initiation Fund Project(BSKY202305), and China Scholarship Council (CSC) - Swansea University scholarship.

\bibliography{SNN}


\section*{Data sets}\label{datasets}
The detailed algorithms of how to generate raw data and implement machine learning are shown in the GitHub link {https://github.com/ChuckShen/DPSNN-code}.


\end{document}